\documentclass[aps,preprint,floatfix,nofootinbib,showpacs]{revtex4-1}
\pdfoutput=1
\usepackage{dcolumn}
\usepackage{bm}
\usepackage{graphicx}
\usepackage{amssymb,amsmath}
\usepackage{multirow}
\usepackage{units,changes}
\usepackage{color,url}
\usepackage{tabu}
\usepackage{array}
\usepackage[colorlinks=true,urlcolor=blue,anchorcolor=blue
,citecolor=blue,filecolor=blue,linkcolor=blue,menucolor=blue
,linktocpage=true,pdfproducer=medialab,pdfa=true]{hyperref}
\usepackage{colordvi}
\usepackage{subcaption}
\def\beqa{\begin{eqnarray}}
\def\eeqa{\end{eqnarray}}

\usepackage{soul}
\usepackage{hhline} 

\begin{document}

\title{Exploring muonphilic ALPs at muon colliders}
\def\slash#1{#1\!\!\!/}

\author{Chih-Ting Lu$^1$, Xiaoyi Luo$^1$, Xinqi Wei$^1$}
\affiliation{
 $^1$ Department of Physics and Institute of Theoretical Physics, Nanjing Normal University, Nanjing, 210023, China \\
}
\date{\today}

\begin{abstract}
Axion-like particles (ALPs) are new particles that extend beyond the standard model (SM) and are highly motivated. When considering ALPs within an effective field theory framework, their couplings with SM particles can be studied independently. It is a daunting task to search for GeV-scale ALPs coupled to muons in collider experiments because their coupling is proportional to the muon mass. However, a recent study by Altmannshofer, Dror, and Gori (2022) highlighted the importance of a four-point interaction, $W$-$\mu$-$\nu_{\mu}$-$a$, as well as interactions from the chiral anomaly which couplings are not dependent on the muon mass. These interactions provide a new opportunity to explore muonphilic ALPs ($\mu$ALPs) at the GeV scale. We have explored various $\mu$ALPs production channels at muon colliders with $\mu$ALPs decaying into a pair of muons. 
Especially, we found a pair of neutrinos accompanied by a $\mu$ALP is a most effective channel to search for $\mu$ALPs in the electrowek violating (EWV) scenario. In contract, a photon plus a $\mu$ALP becomes a better channel to search for $\mu$ALPs in the electroweak preserving (EWP) scenario because there is no $W$-$\mu$-$\nu_{\mu}$-$a$ interaction in this situation. Most importantly, we found that the future bounds for $\mu$ALPs in EWV scenario are much stronger than the ones in EWP scenario and the existing bounds for exploring $\mu$ALPs with $1$ GeV $\leq m_a\lesssim M_W$. 
\end{abstract}

\maketitle

\section{Introduction}
Axion-like particles (ALPs) are predicted to exist in a wide range of models that extend beyond the standard model (SM). The QCD axion, introduced originally to solve the strong CP problem, is one such model~\cite{Peccei:1977hh,Weinberg:1977ma,Wilczek:1977pj,Kim:1979if,Kim:2008hd}. ALPs can also be generated from different spontaneous symmetry breaking patterns of global symmetries~\cite{Preskill:1982cy,Abbott:1982af,Dine:1982ah,Bagger:1994hh} as well as in string theory~\cite{Svrcek:2006yi,Arvanitaki:2009fg,Cicoli:2012sz,Visinelli:2018utg} and models of extra dimensions~\cite{Chang:1999si,Bastero-Gil:2021oky}. The broad spectrum of possible ALP masses makes them an attractive candidate for a variety of astrophysical and cosmological phenomena~\cite{Jaeckel:2010ni}. Sub-eV ALPs have been proposed as potential candidates for dark matter~\cite{Arias:2012az}. ALPs at different mass scales can also serve other purposes, such as acting as mediators to the dark sector~\cite{Zhevlakov:2022vio,Bharucha:2022lty}, influencing the structure of the electroweak phase transition~\cite{Jeong:2018jqe,Im:2021xoy}, and offering solutions to the hierarchy problem of the Higgs boson mass~\cite{Graham:2015cka}. 
Understanding the characteristics and roles of ALPs is essential for unraveling the mysteries of the universe and advancing our knowledge of particle physics.

Various methods have been developed to search for ALPs, including laboratory-based experiments~\cite{Sikivie:1983ip}, astrophysical observations~\cite{CAST:2017uph}, and searches for ALPs in high-energy collisions~\cite{Bauer:2018uxu}. The current constraints on ALPs rely on their coupling strength and mass. For example, astrophysical observations of the diffuse gamma-ray background provide tight constraints on the coupling strength of sub-eV ALPs to photons~\cite{Li:2022jgi,Eckner:2022rwf,Mastrototaro:2022kpt,Li:2022pqa}, while experiments based on the LEP and LHC can limit the coupling strength of high-mass ALPs to SM particles~\cite{OPAL:2002vhf,Mimasu:2014nea,Jaeckel:2015jla,ATLAS:2014jdv,ATLAS:2015rsn,Knapen:2016moh}. With the advancements of experimental techniques, these bounds are expected to become even more stringent in the future, offering exciting new prospects of investigating the properties of ALPs.

In this work, we focus on studying muonphilic ALPs ($\mu$ALPs), a specific type of ALP that predominantly interacts with muons~\cite{Bollig:2020xdr,Croon:2020lrf,Buen-Abad:2021fwq,Ge:2021cjz,Caputo:2021rux,Cheung:2022umw,Liu:2022tqn,Calibbi:2022izs}. These ALPs can be considered in an effective field theory framework~\cite{Brivio:2017ije,Bauer:2017ris,Bauer:2018uxu,Ebadi:2019gij,Bauer:2020jbp,Bauer:2021mvw}, allowing us to study their couplings with SM particles independently. Bounds on $\mu$ALPs for $m_a < 2m_{\mu}$ have already been obtained from searches in supernovae~\cite{Bollig:2020xdr, Croon:2020lrf,Caputo:2021rux} and atmospheric air showers~\cite{Cheung:2022umw}. For $2m_{\mu} < m_a \lesssim {\cal O}(1)$ GeV, $\mu$ALPs can be largely produced in fixed target experiments~\cite{Zhevlakov:2022vio}, low-energy $e^{+}e^{-}$ colliders~\cite{BaBar:2016sci}, and Tera Z factories~\cite{Calibbi:2022izs}. However, searching for GeV-scale $\mu$ALPs at high-energy colliders is challenging due to the small $\mu$ALP production rate, as the coupling is proportional to the muon mass. Therefore, proposing new $\mu$ALP production channels with sufficiently large cross sections at high-energy colliders is crucial to search for GeV-scale $\mu$ALPs.

Recently, a four-point interaction ($W$-$\ell$-$\nu$-$a$), which has a coupling that is independent of the charged lepton mass, has been proposed for the search of leptophilic ALPs~\cite{Altmannshofer:2022izm}. This interaction is expected to arise from decays of $\pi^{\pm}$, $K^{\pm}$ mesons, and the $W$ boson, with the novel energy enhancement effect. Similarly, this kind of $W$-$\ell$-$\nu$-$a$ interaction with energy enhancement effect has also been proposed as a promising approach for the search of leptophilic ALPs via t-channel processes ($\ell^+\ell^-\rightarrow\overline{\nu_{\ell}}a\nu_{\ell}$ and $\ell^- p\rightarrow\nu_{\ell}aj$) at high-energy colliders~\cite{Lu:2022zbe}. In this study, we investigated the production of GeV-scale $\mu$ALPs from the above t-channel processes and their decay into a pair of muons at muon colliders~\cite{AlAli:2021let,MuonCollider:2022xlm,Black:2022cth}. Notably, when a light $\mu$ALP is highly-boosted produced, the resulting pair of muons from the $\mu$ALP decay is too collimated to pass the muon isolation criteria, and forms a novel object known as a muon-jet~\cite{Arkani-Hamed:2008kxc,Baumgart:2009tn,Bai:2009it,Cheung:2009su,Falkowski:2010cm,Han:2015lma,Izaguirre:2015pga,Izaguirre:2015zva,Chang:2016lfq,Kim:2016fdv,Dube:2017jgo,Zhang:2021orr}.

We investigate three major signal processes at muon colliders : $\mu^+\mu^-\rightarrow\nu_{\mu}a\overline{\nu_{\mu}}$, $\mu^{+}\mu^{-}\rightarrow \gamma a$ and $\mu^{+}\mu^{-}\rightarrow \mu^{+}\mu^{-} a$. These signal production modes mainly rely on a four-point interaction, $W$-$\mu$-$\nu_{\mu}$-$a$, and/or interactions from the chiral anomaly which couplings are not dependent on the muon mass. Generally, $\mu^+\mu^-\rightarrow\nu_{\mu}a\overline{\nu_{\mu}}$ yields the largest cross section, followed by $\mu^{+}\mu^{-}\rightarrow \gamma a$ and $\mu^{+}\mu^{-}\rightarrow \mu^{+}\mu^{-} a$ in the electroweak violating (EWV) scenario. However, there is no $W$-$\mu$-$\nu_{\mu}$-$a$ interaction in the electroweak preserving (EWP) scenario, and therefore,  $\mu^{+}\mu^{-}\rightarrow \gamma a$ yields the largest cross section. In the \textbf{EWV} scenario, we discovered that the channel $\mu^+\mu^-\rightarrow\nu_{\mu}a\overline{\nu_{\mu}}$ with the $W$-$\ell$-$\nu$-$a$ interaction is the most important one among these channels because of its novel energy-enhancement behavior. Our findings suggest that searching for the signature of two isolated muons (or a muon-jet) plus missing energy in the \textbf{EWV} scenario at muon colliders can provide much stronger bounds than existing ones. On the other hand, searching for the signatures of two isolated muons (or a muon-jet) plus a photon and four isolated muons (or a muon-jet plus two isolated muons) in the \textbf{EWP} scenario at muon colliders may only slightly exceed existing bounds. Therefore, the muon collider is an ideal machine to search for $\mu$ALPs and it can also explore a $\mu$ALP belonging to the \textbf{EWV} or \textbf{EWP} scenario.

The plan of this paper is as follows. In Sec.~\ref{sec:model}, we provide a brief review of ALP-muon interactions and $\mu$ALP decay modes. The method to distinguish different ALP-muon interaction types using $\mu^+\mu^-\rightarrow\nu_{\mu}a\overline{\nu_{\mu}}$, $\mu^{+}\mu^{-}\rightarrow \gamma a$ and $\mu^{+}\mu^{-}\rightarrow \mu^{+}\mu^{-} a$ processes is discussed in Sec.~\ref{sec:Xsec}. We present the results of a full signal-to-background analysis at muon colliders and compare them with existing bounds of the $\mu$ALP in Sec.~\ref{sec:analysis}. Finally, we summarize our findings in Sec.~\ref{sec:final}. Supplementary materials, including kinematic distributions for both signals and SM backgrounds and other tables are provided in Appendix~\ref{app:rec}.

\section{Review on ALP-muon interactions}
\label{sec:model}

 We consider ALPs, generated from the global Peccei-Quinn (PQ) symmetry~\cite{Peccei:1977hh}, $U(1)_{\text{PQ}}$, breaking. Based on the structure of the PQ symmetry, $a(x)\rightarrow a(x)+\text{const}$, the Lagrangian can be written in the form  
${\cal L}_{\mu\text{ALP}} = \partial_{\nu}a ~J^{\nu}_{ \text{PQ},\mu}$.  
The general muon current is in the form, 
\begin{equation} 
\label{eq:Jint}
J^{\nu}_{\text{PQ},\mu} = \frac{c^V_{\mu}}{2\Lambda}\overline{\mu}\gamma^{\nu}\mu + \frac{c^A_{\mu}}{2\Lambda}\overline{\mu}\gamma^{\nu}\gamma_5\mu + \frac{c_{\nu_{\mu}}}{2\Lambda}\overline{\nu_{\mu}}\gamma^{\nu} P_L \nu_{\mu} \,, 
\end{equation} 
where $\Lambda$ is the new physics scale, and $c^V_{\mu}$, $c^A_{\mu}$, $c_{\nu_{\mu}}$ are dimensionless couplings. Without the assumption of electroweak invariance, the condition $c_{\nu_{\mu}} = c^V_{\mu} -c^A_{\mu}$ in Eq.~(\ref{eq:Jint}) can be released\footnote{Note the dimensional five operators with electroweak invariance to generate the first and the third terms in Eq.~(\ref{eq:Jint}) are discussed in Ref.~\cite{Altmannshofer:2022izm}.}. 
After integrating by parts of this Lagrangian, the ${\cal L}_{\mu\text{ALP}}$ can be written as~\cite{Altmannshofer:2022izm}  
\begin{align} 
a ~\partial_{\nu}J^{\nu}_{\text{PQ},\mu} = & ~i c^A_{\mu}\frac{m_{\mu}}{\Lambda}~a\overline{\mu}\gamma_5\mu + \frac{\alpha_{\text{em}}}{4\pi\Lambda} \bigg[  \frac{ c^V_{\mu} -c^A_{\mu} + c_{\nu_{\mu}}}{4 s^2 _W}~a W^{+}_{\mu\nu}\tilde W ^{-,\mu\nu} \notag \\ 
& + \frac{c^V_{\mu} - c^A_{\mu} (1 -4 s^2_W)}{2s _W c_W}~a F_{\mu\nu}\tilde{Z}^{\mu\nu} - c^A_{\mu}~a F_{\mu\nu} \tilde{F}^{\mu\nu} + \notag \\ 
& \frac{c^V_{\mu} (1 -4 s^2_W) -c^A_{\mu} (1 -4 s^2_W +8 s^4_W)  + c_{\nu_{\mu}}}{8 s^2_W c^2_W}~a Z_{\mu\nu}\tilde{Z}^{\mu\nu}\bigg]  \notag \\ 
& + \frac{ig_W}{2\sqrt{2}\Lambda}(c^A_{\mu} - c^V_{\mu} + c_{\nu_{\mu}})~a (\bar\mu \gamma^{\nu} P _L \nu_{\mu}) W_{\nu}^{-} ~+~\text{h.c.}  \,, \label{eq:int} 
\end{align} 
the symbols $W^{\pm}_{\mu\nu}$, $Z_{\mu\nu}$, $F_{\mu\nu}$ represent the field strength tensors of massive gauge bosons $W^{\pm}$, $Z$ and the massless photon, and the dual field strength tensor is defined as $\tilde{F}_{\mu\nu}=\frac{1}{2}\epsilon_{\mu\nu\rho\sigma}F^{\rho\sigma}$. On the other hand, $\alpha_{\text{em}}$ is the fine structure constant, $g_W$ is the weak coupling constant and $s_W$ and $c_W$ are the sine and cosine of the weak mixing angle, respectively.

In Eq.~(\ref{eq:int}), we label the first term as "$\boldsymbol{a\mu\mu}$", which can generate $\mu$ALPs through the muon radiation. However, this term is suppressed by $m_{\mu}/\Lambda$, necessitating high-intensity experiments to search for light $\mu$ALPs. The second to the fourth terms, labeled as "$\boldsymbol{aVV'}$", arise from the chiral anomaly and can produce light $\mu$ALPs through flavor-changing processes in meson decays~\cite{Izaguirre:2016dfi,Gori:2020xvq,Bauer:2021mvw}. Heavier $\mu$ALPs can also be produced from these terms through gauge boson fusion and associated gauge boson production processes, despite not being proportional to $m_{\mu}$, but having a $\alpha_{\text{em}}/4\pi$ suppression. The terms in the final line of Eq.~(\ref{eq:int}), labeled as "$\boldsymbol{aW\mu\nu}$", are often overlooked in the literature~\cite{Raffelt:1987yt}. However, they are critical to our work, particularly for searching for $\mu$ALPs in the GeV scale. This four-point interaction, $W$-$\mu$-$\nu_{\mu}$-$a$, vanishes when the general muon current in Eq.~(\ref{eq:Jint}) respects the electroweak symmetry. Moreover, this interaction is not related to $m_{\mu}$ and has an obvious $\left(\text{energy}/\Lambda\right)$ enhancement in specific processes. This enhancement behavior is crucial in constraining light $\mu$ALPs through decays of the $W$ boson and charged mesons~\cite{Altmannshofer:2022izm}, as well as in searching for heavier $\mu$ALPs in $t$-channel processes such as $\mu^{+}\mu^{-}\rightarrow\nu_{\mu} a\overline{\nu_{\mu}}$ at muon colliders under the EWV scenario which will be defined in the next section.

On the other hand, searching for $\mu$ALPs in collider experiments will depend on their decay modes. For $\mu$ALP masses below the electroweak scale ($m_a\lesssim M_W$), their dominant decay modes are to $\mu^{+}\mu^{-}$ and $\gamma\gamma$~\cite{Bauer:2017ris,Bauer:2018uxu,Chang:2021myh}. The decay widths are given by
\begin{equation}
\Gamma_{a\rightarrow\mu^{+}\mu^{-}} = \frac{(c^A_{\mu})^2 m^2_{\mu} m_a}{8\pi\Lambda^2}\sqrt{1-\frac{4m^2_{\mu}}{m^2_a}} ~,~~
\Gamma_{a\rightarrow\gamma\gamma} = \frac{g^2_{a\gamma\gamma}m^3_a}{64\pi} ,
\end{equation}
where the coupling constant $g_{a\gamma\gamma}$ is determined by the chiral anomaly and one-loop triangle Feynman diagrams, and can be expressed as 
\begin{equation}
g_{a\gamma\gamma} = \frac{\alpha_{\text{em}}}{\pi}\frac{c^A_{\mu}}{\Lambda}\lvert 1 - {\cal F} (\frac{m^2_a}{4m^2_{\mu}})\rvert
\end{equation} 
and the loop function ${\cal F} (z > 1) = \frac{1}{z}\text{arctan}^2\left(\frac{1}{\sqrt{1/z -1}}\right)$. Here, we only consider the contribution from the muon loop, as the contribution from the $W$ boson is strongly suppressed and can be safely neglected. 

\begin{figure}[tb]
\centering{\includegraphics[width=0.6\textwidth]{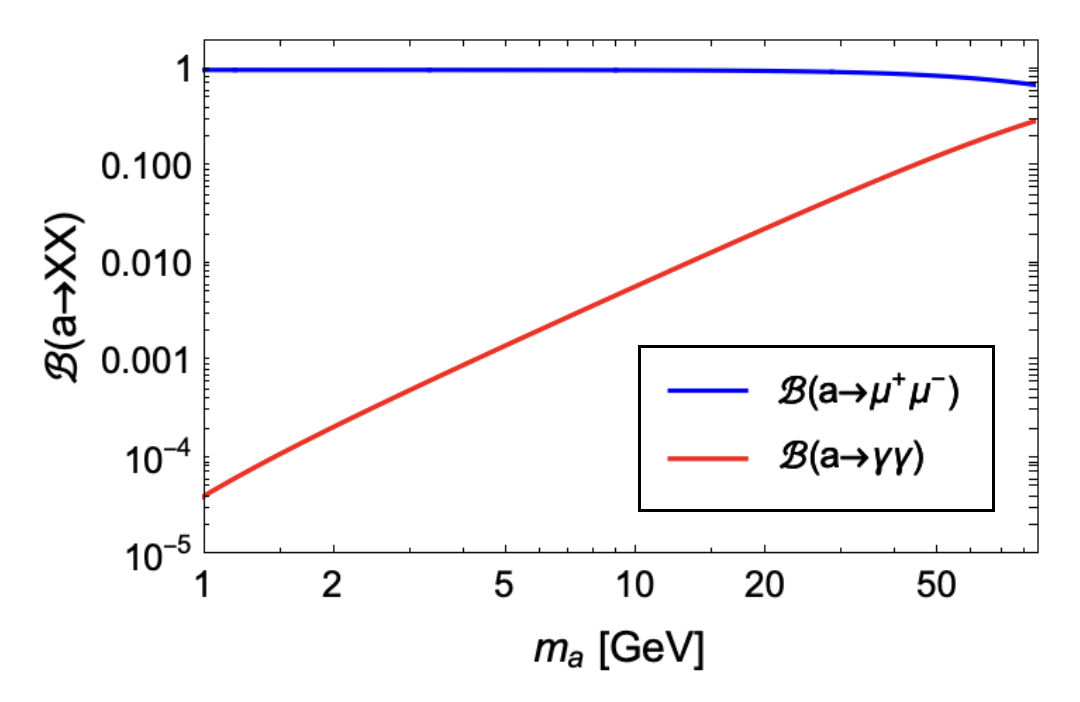}}
\caption{The decay branching ratios of $\mu$ALP below the electroweak scale ($m_a\lesssim M_W$). 
}
\label{fig:lALP_BR}
\end{figure}

The Fig.~\ref{fig:lALP_BR} shows the branching ratios for $a\rightarrow \mu^{+}\mu^{-}$ and $a\rightarrow\gamma\gamma$. When $m_a\lesssim M_W$, the dominant decay mode of $\mu$ALP is $a\rightarrow \mu^{+}\mu^{-}$. Since the partial decay width of $a\rightarrow\gamma\gamma$ depends slightly on the muon mass and scales with $m_a^3$, we can expect the branching ratio of $a\rightarrow\gamma\gamma$ to increase with the $\mu$ALP mass. It is important to note that this result is opposite to that of the electrophilic ALP in Ref.~\cite{Lu:2022zbe} because the muon mass is much larger than the electron mass.

\section{Distinguish different ALP-muon interaction types at muon colliders}
\label{sec:Xsec}

\begin{figure}[tb]
\centering{\includegraphics[width=0.7\textwidth]{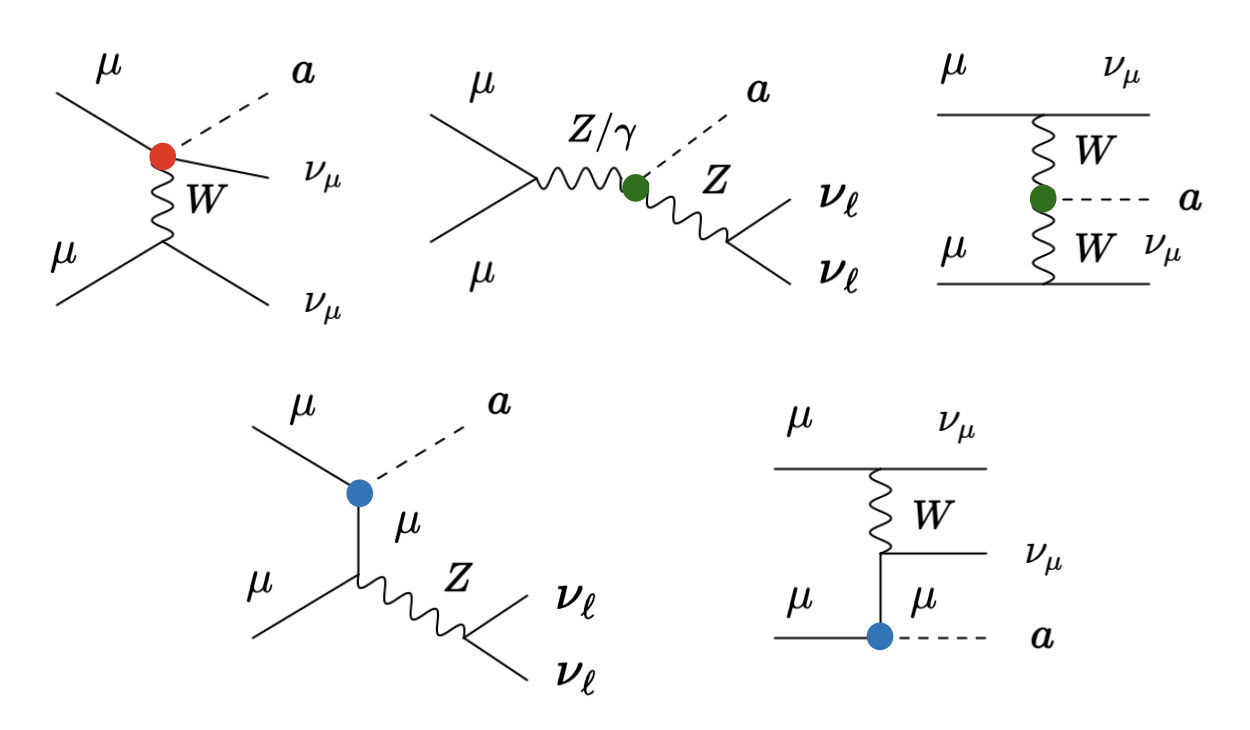}}
\caption{Feynman diagrams for $\mu^{+}\mu^{-}\rightarrow\nu_{\mu} a\overline{\nu_{\mu}}$. Here the color markers indicate red for $\boldsymbol{aW\mu\nu}$ interaction, green for $\boldsymbol{aVV'}$ interaction and blue for $\boldsymbol{a\mu\mu}$ interaction.  
}
\label{fig:Feyn1}
\end{figure}

\begin{figure}[tb]
\centering{\includegraphics[width=0.6\textwidth]{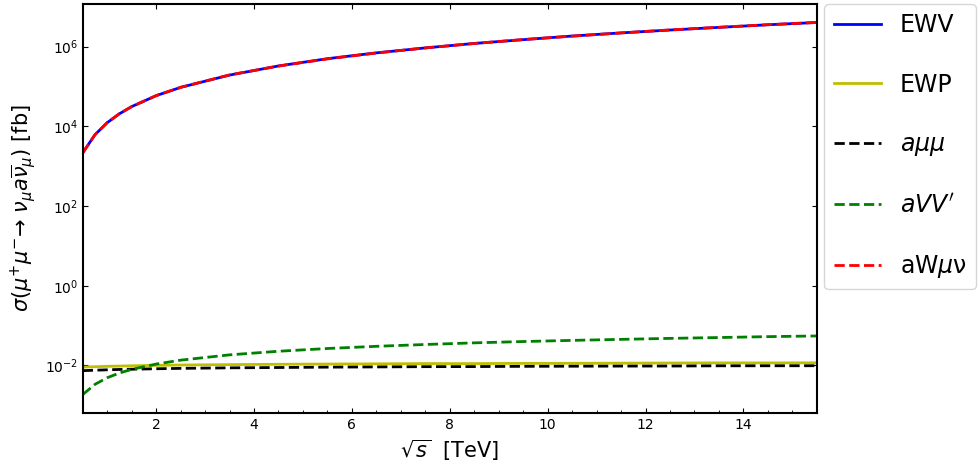}}
\caption{The energy enhancement behavior of cross sections in $\mu^{+}\mu^{-}\rightarrow\nu_{\mu} a\overline{\nu_{\mu}}$ with $m_a = 10$ GeV, $c^A_{\mu}/\Lambda = 10$ TeV$^{-1}$, $c^V_{\mu}=c_{\nu_{\mu}}=0$ (\textbf{EWV}: solid-blue line) and $c^A_{\mu} / \Lambda = c^V_{\mu}/ \Lambda = 10$ TeV$^{-1}$, $c_{\nu_{\mu}}=0$ (\textbf{EWP}: solid-olive line). In the \textbf{EWV} scenario, the dashed-red, dashed-green, and dashed-black lines are labeled as contributions from $\boldsymbol{aW\mu\nu}$, $\boldsymbol{aVV'}$ and $\boldsymbol{a\mu\mu}$ interactions, respectively. 
} 
\label{fig:ALP_Xsec}
\end{figure}

In this section, we focus on distinguishing between different types of ALP-muon interactions at muon colliders. First, we consider the signal process $\mu^{+}\mu^{-}\rightarrow\nu_{\mu}a\overline{\nu_{\mu}}$ with the relevant Feynman diagrams showing in Fig.~\ref{fig:Feyn1} and numerically investigate the energy enhancement behavior of this process at muon colliders. To implement ${\cal L}_{\mu\text{ALP}}$ from Eq.~(\ref{eq:int}), we use FeynRules~\cite{Alloul:2013bka} and calculate cross sections for this process using Madgraph5\underline{\hspace{0.5em}}aMC@NLO~\cite{Alwall:2014hca}, while varying the center-of-mass energy. As we know, the condition $c_{\nu_{\mu}} = c^V_{\mu} -c^A_{\mu}$ is a criterion to determine whether the ALP effective field theory is electroweak invariant or not. Therefore, we set $c^A_{\mu} / \Lambda = 10$ TeV$^{-1}$ and $c^V_{\mu} = c_{\nu_{\mu}}=0$ as a benchmark point for the electroweak violating~(\textbf{EWV}) scenario. Similarly, we set $c^A_{\mu}/ \Lambda = c^V_{\mu}/ \Lambda = 10$ TeV$^{-1}$ and $c_{\nu_{\mu}}=0$ as a benchmark point for the electroweak preserving~(\textbf{EWP}) scenario. We vary the center-of-mass energy $\sqrt{s}$ between $1-15$ TeV with $m_a = 10$ GeV at muon colliders. Fig.~\ref{fig:ALP_Xsec} shows the energy enhancement behavior of cross section in $\mu^{+}\mu^{-}\rightarrow\nu_{\mu} a\overline{\nu_{\mu}}$, where the full contributions from the \textbf{EWV} and \textbf{EWP} scenarios are depicted in solid lines, and the contributions from $\boldsymbol{aW\mu\nu}$, $\boldsymbol{aVV'}$, and $\boldsymbol{aW\mu\nu}$ in the \textbf{EWV} scenario are depicted in dashed lines.

As shown in Fig.~\ref{fig:ALP_Xsec}, the leading contribution in the \textbf{EWV} scenario comes from the $\boldsymbol{aW\mu\nu}$ interaction, with the subleading contribution from $\boldsymbol{aVV'}$ interaction. The contribution from $\boldsymbol{aVV'}$ interaction is about seven orders of magnitude smaller than that from $\boldsymbol{aW\mu\nu}$, as depicted in dashed lines in Fig.~\ref{fig:ALP_Xsec}. For $\sqrt{s} = 1-2$ TeV, the energy enhancement behaviors from these two interactions are evident because the momentum transfer size becomes large enough, making $\boldsymbol{aW\mu\nu}$ and $\boldsymbol{aVV'}$ interactions important. However, as energy continues to increase, the growth rate becomes gentler because these two leading contributions steadily increase with the center-of-mass energy as $\left(\text{energy}/\Lambda\right)$.

Our numerical analysis reveals that the contribution from the $\boldsymbol{aW\mu\nu}$ interaction is much greater than those from the $\boldsymbol{aVV'}$ and $\boldsymbol{a\mu\mu}$ interactions because of the novel energy enhancement behavior. Therefore, we show the analytical form for the amplitude square with the average (sum) over initial (final) polarization for the $\boldsymbol{aW\mu\nu}$ interaction in the process $\mu^+(p_1)\mu^-(p_2)\rightarrow\nu_{\mu}(q_1)a(q_2)\overline{\nu_{\mu}}(q_3)$,    
\begin{align} 
\overline{\lvert{\cal M}\rvert ^2} = & \frac{g^4_W\left( c^A_{\mu}-c^V_{\mu}+c_{\nu_{\mu}}\right) ^2}{32\Lambda^2}\left(\frac{1}{k^2-M^2_W}+\frac{1}{k^{\prime 2}-M^2_W}\right) ^2 \notag \\ \notag
& \times\left( s-2m^2_{\mu}\right)\left[ s-m^2_a -2q_2\cdot (q_1 +q_3)\right] \notag  \,, 
\label{eq:amplitude} 
\end{align} 
where $s = \left( p_1 +p_2\right) ^2 = \left( q_1 + q_2 + q_3\right) ^2$, $k = p_2 -q_3$ and $k' = p_1 -q_1$. It shows that the amplitude square can be enhanced when the momentum transfer in the $t$-channel process is large enough.

In the \textbf{EWV} scenario, the contribution from $\boldsymbol{a\mu\mu}$ interaction is negligible, while in the \textbf{EWP} scenario, there are both $\boldsymbol{aVV'}$ and $\boldsymbol{a\mu\mu}$ interactions in $\mu^{+}\mu^{-}\rightarrow\nu_\mu a\overline{\nu_\mu}$. The cross sections have no obvious change with the center-of-mass energy increase in the \textbf{EWP} scenario due to the lack of energy enhancement effect. 
Lastly, the cross sections in the \textbf{EWV} scenario are more than six orders of magnitude larger than those in the \textbf{EWP} scenario for $\mu^{+}\mu^{-}\rightarrow\nu_{\mu} a\overline{\nu_{\mu}}$ process in Fig.~\ref{fig:ALP_Xsec}. 
This is because there is $\boldsymbol{aW\mu\nu}$ interaction in the \textbf{EWV} scenario, but not in the \textbf{EWP} scenario, and this interaction contributes to almost the entire cross-section amount in the \textbf{EWV} scenario. Therefore, this process is powerful to distinguish $\mu$ALPs in the \textbf{EWV} scenario from the \textbf{EWP} scenario. 

\begin{table}[ht!]
\begin{center}\begin{tabular}{|c|c|c|}
\hline \multirow{2}{*}{production channel} & \multicolumn{2}{|c|}{cross section [fb]} \\
\cline{2-3}  & ~$\textbf{EWV}$~ &  ~$\textbf{EWP}$~ \\ 
\hline $\mu^{+}\mu^{-}\rightarrow\nu_{\mu} a \overline{\nu}_{\mu}$ & $3.13\times 10^{4}$ & $9.69\times 10^{-3}$  \\
\hline $\mu^{+}\mu^{-}$ $\rightarrow$ $\mu^{+}\mu^{-}a$ & $1.45\times 10^{-2}$ & $1.69\times 10^{-2}$  \\
\hline $\mu^{+}\mu^{-}$ $\rightarrow$ $a \gamma$ & $7.72\times 10^{-2}$ & $8.18\times 10^{-2}$  \\
\hline $\mu^{+}\mu^{-}$ $\rightarrow$ $Z a$ & $3.58\times 10^{-3}$ & $2.54\times 10^{-2}$ \\ 
\hline \end{tabular} \caption{The cross sections of different $\mu$ALP production channels at a muon collider with $\sqrt{s} = 3$ TeV has been shown in this table with the benchmark point $m_a = 10$ GeV and $c^A_{\mu} / \Lambda = 10$ TeV$^{-1}$.}
\label{tab:channel_1}
\end{center}
\end{table}

Next, we discuss four optimal channels for searching for $\mu$ALPs at a muon collider. Among these channels, we specifically consider those $\mu$ALP couplings which are independent of the muon mass. These $\mu$ALP production channels are $\mu^+\mu^-\rightarrow\nu_{\mu}a\overline{\nu_{\mu}}$, $\mu^{+}\mu^{-}\rightarrow \gamma a$, $\mu^{+}\mu^{-}\rightarrow \mu^{+}\mu^{-} a$ and $\mu^{+}\mu^{-}\rightarrow Z a$. To compare these $\mu$ALP production channels at a muon collider with $\sqrt{s} = 3$ TeV, we calculated their cross sections in both \textbf{EWV} and \textbf{EWP} scenarios with the benchmark point $m_a = 10$ GeV and $c^A_{\mu} / \Lambda = 10$ TeV$^{-1}$, as shown in Table~\ref{tab:channel_1}. In the \textbf{EWV} scenario, we found that the $\mu^+\mu^-\rightarrow\nu_{\mu}a\overline{\nu_{\mu}}$ channel has the largest cross section, due to its energy-enhancing behavior caused by the $W$-$\mu$-$\nu_{\mu}$-$a$ interaction. The cross section of this channel is about six to seven orders of magnitude higher than that of other channels. However, in the \textbf{EWP} scenario, there is no energy-enhancing behavior in the $\mu^+\mu^-\rightarrow\nu_{\mu}a\overline{\nu_{\mu}}$ channel such that the cross section of this channel becomes smaller than that of other channels. At this point, the cross section of the $\mu^{+}\mu^{-}\rightarrow \gamma a$ channel is the largest, making it the most prospective search channel in the \textbf{EWP} scenario.

\section{Signal-to-background analysis at a muon collider}
\label{sec:analysis}

In this section, we investigate signal and background analysis on three specific search channels: $\mu^+\mu^-\rightarrow\nu_{\mu}a\overline{\nu}_{\mu}$, $\mu^{+}\mu^{-}\rightarrow \gamma a$ and $\mu^{+}\mu^{-}\rightarrow \mu^{+}\mu^{-} a$ processes. 
Our goal is to predict the future bounds for GeV-scale $\mu$ALPs at a muon collider and compare them with existing bounds.

\subsection{Exploring $\mu^{+}\mu^{-}\rightarrow \nu_{\mu}a\overline{\nu}_{\mu}$ in the EWV scenario} 
\label{sec:uuvv}

\begin{figure}[tb]
\centering{\includegraphics[width=0.48\textwidth]{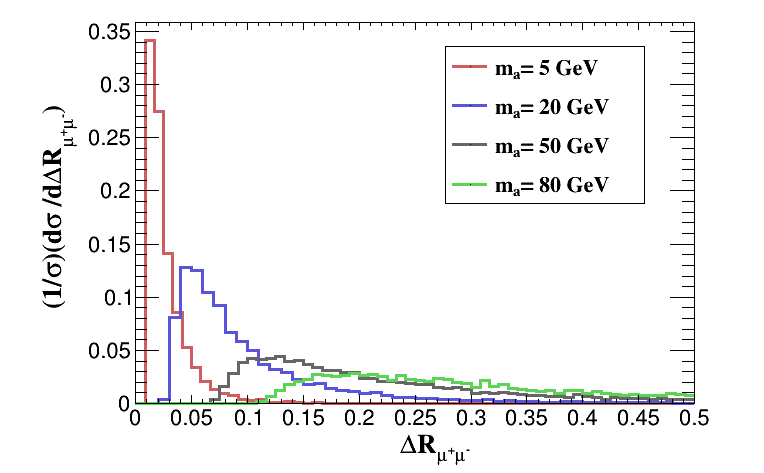}}
\centering{\includegraphics[width=0.48\textwidth]{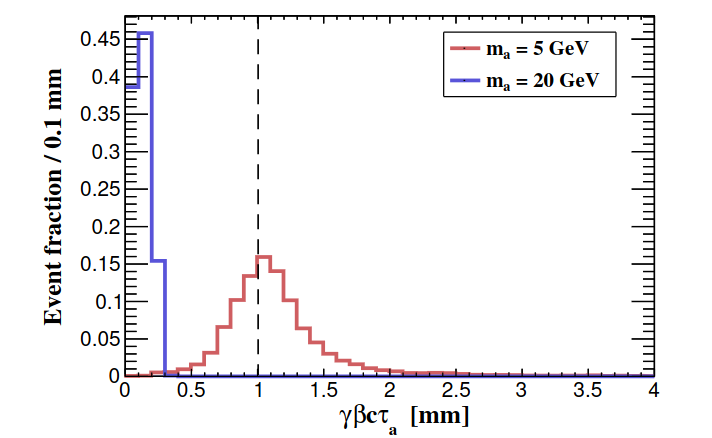}}
\caption{Left panel : Distribution of the opening angle between two muons, $\Delta$ $R_{\mu^{+}\mu^{-}}$, from $\mu^+\mu^-\rightarrow\nu_{\mu}a\overline{\nu_{\mu}}$ at the muon collider with $\sqrt{s}=3$ TeV. Four benchmark mass values of $\mu$ALP, $m_{a}=5$, $20$, $50$, $80$ GeV are displayed. Right panel : Distribution of the $\mu$ALP lab frame decay length from $\mu^+\mu^-\rightarrow\nu_{\mu}a\overline{\nu_{\mu}}$ at the muon collider with $\sqrt{s}=3$ TeV. The benchmark points $m_a =5$, $20$ GeV with $c^A_{\mu} / \Lambda = 0.1$ TeV$^{-1}$ are considered.
}
\label{fig:DR}
\label{fig:L}
\end{figure}

As an illustration, we analyze the process $\mu^+\mu^-\rightarrow\nu_{\mu}a\overline{\nu}_{\mu}$ in the \textbf{EWV} scenario and its relevant SM backgrounds in the context of the popular muon collider proposal with $\sqrt{s}=3$ TeV~\cite{MuonCollider:2022xlm,Black:2022cth}. According to Fig.~\ref{fig:lALP_BR}, the $\mu$ALP mainly decays to $\mu^+\mu^-$ when $m_a\lesssim M_W$. Hence, we focus on the $a\rightarrow\mu^+\mu^-$ decay mode in our analysis. The $\mu$ALP becomes highly boosted at the muon collider when it is light enough, so $\mu^{+} \mu^{-}$ in the final state may be too collimated to pass the muon isolation criterion at detectors. Taking a cone size $R=0.1$ as the muon isolation criterion at the muon collider, we find a pair of muons cannot be isolated to each other at detectors when $m_a \lesssim 15$ GeV (parton-level) in the left panel of Fig.~\ref{fig:DR}. We can group this kind of collimated, non-isolated muons as a special signature "muon-jet" ($J_{\mu}$) which is a non-QCD jet-like structure and deposits most of its energy in the the muon spectrometer and has distinct signature from QCD jets. Therefore, we classify the signal signatures into two categories: (1) two isolated muons plus missing energy (${\:/\!\!\!\! E}$) for $m_a\gtrsim 15$ GeV, and (2) a $J_{\mu}$ plus ${\:/\!\!\!\! E}$ for $m_a \lesssim 15$ GeV. 

\begin{table}[ht!]
	\begin{center}\begin{tabular}{|c|c|c|c|}\hline cut flow in $\sigma$ [fb] & ~signal~ & ~$\nu_{\ell}\overline{\nu_{\ell}}\mu^{+}\mu^{-}$~ 
			
			& ~$t\overline{t}$~ \\ 
			\hline Generator & $2.54$ & $162.70$ & $4.15\times 10^{-2}$ \\
			\hline cut-(1) & $1.78$ & $18.60$ & $7.94\times 10^{-3}$ \\
			\hline cut-(2) & $1.78$ & $11.28$ & $7.21\times 10^{-3}$ \\
			\hline cut-(3) & $1.78$ & $11.27$ & $3.54\times 10^{-4}$ \\
			\hline cut-(4) & $1.74$ & $0.15$ & $2.12\times 10^{-5}$ \\ 
			\hline cut-(5) & $1.47$ & $1.23\times 10^{-2}$ & $8.30\times 10^{-7}$ \\
            \hline cut-(6) & $1.35$ & $6.17\times 10^{-3}$ & $0$ \\
			\hline \end{tabular} \caption{The cut-flow table for $\mu^{+}\mu^{-}\rightarrow\nu_{\ell} (a\rightarrow\mu^{+}\mu^{-})\overline{\nu_{\ell}}$ and relevant SM backgrounds with signature of two isolated muons plus ${\:/\!\!\!\! E}$. The benchmark point $m_a = 50$ GeV with $c^A_{\mu} / \Lambda = 0.1$ TeV$^{-1}$ for signal is chosen. Each event selection has been mentioned in the main text. The "Generator" means the cross sections in parton-level calculated by Madgraph5\underline{\hspace{0.5em}}aMC@NLO.}
		\label{tab:CF_uu_1}
	\end{center}
\end{table} 

For the first signal signature, two relevant SM backgrounds : $\mu^{+}\mu^{-}\rightarrow \nu_{\ell}\overline{\nu_{\ell}}\mu^{+}\mu^{-}$ and $\mu^{+}\mu^{-}\rightarrow t\overline{t}\rightarrow (b\mu^+\nu_{\mu})(\overline{b}\mu^-\overline{\nu_{\mu}})$ are considered. We choose the benchmark point $m_a = 50$ GeV with $c^A_{\mu} / \Lambda = 0.1$ TeV$^{-1}$ to display the signal features. To generate Monte Carlo samples for both signal and background processes, we use Madgraph5\underline{\hspace{0.5em}}aMC@NLO~\cite{Alwall:2014hca} and pass them to Pythia8~\cite{Sjostrand:2007gs} for QED and QCD showering and hadronization effects. We impose pre-selection cuts ($P^{\mu}_T > 5$ GeV and $\lvert\eta_{\mu}\rvert < 2.5$) at the parton-level for both the signal and backgrounds. To simulate the detector effects, we use the muon collider template in Delphes3~\cite{deFavereau:2013fsa} which the muon isolation criterion is consistent with Ref.~\cite{Yang:2021zak,Haghighat:2021djz}. We use the Cambridge/Aachen ($C/A$) jet clustering algorithm~\cite{Dokshitzer:1997in,Wobisch:1998wt} and consider a b-jet tagging efficiency of $\epsilon_b = 0.8$ with charm-jet and light-jet fake rates of $P_{c\rightarrow b}=0.1$ and $P_{j\rightarrow b}=10^{-3}$, respectively. The following event selections to identify the signal signature and suppress background events are required :
\begin{itemize}
	\item (1) $N(\mu)\ge 2$ with $ P_T^{\mu_1} > 200 $ GeV, $ P_T^{\mu_2} > 10$ GeV, $\lvert\eta_{\mu_{1,2}}\rvert < 1.5$, 
	\item (2) $ 1500 < {\:/\!\!\!\! E} < 2800$ GeV and $\lvert\eta_{{\:/\!\!\!\! E}}\rvert < 1.8$,
	\item (3) Veto $N(b)\ge1$ GeV with $ P_T^{b} > 25 $ GeV,
    \item (4) ${\:/\!\!\!\! E}/M_{\mu_1\mu_2} > 32$,
	\item (5) $\lvert M_{\mu_1\mu_2}-m_a\rvert < 2$ GeV, 
	\item (6) $3.0 < \Delta\phi_{\mu_1, {\:/\!\!\!\! E}} < 3.3$ and $2.9 < \Delta\phi_{\mu_2, {\:/\!\!\!\! E}} < 3.5$, 
\end{itemize}
where $P_T^{\mu_1}$, $P_T^{\mu_2}$ ($\eta_{\mu_1}$, $\eta_{\mu_2}$) are the transverse momentum (pseudorapidity) of leading and subleading energetic muons, ${\:/\!\!\!\! E}$ is the missing energy, $M_{\mu_1\mu_2}$ is the invariant mass of a muon pair, $\Delta\phi_{\mu_i, {\:/\!\!\!\! E}}$ is the azimuthal angle between the i-th muon and ${\:/\!\!\!\! E}$.   
The cut-flow table including signal and backgrounds for each event selection is listed in Table.~\ref{tab:CF_uu_1} and some kinematic distributions are shown in Fig.~\ref{fig:3} of Appendix~\ref{app:rec}.

First, we found two isolated muons and ${\:/\!\!\!\! E}$ in the central region of signal events. To select candidate events, we applied the following trigger criteria : $P_T^{\mu_1} > 200$ GeV, $P_T^{\mu_2} > 10$ GeV, and ${\:/\!\!\!\! E} > 1500$ GeV. In Fig.~\ref{fig:3}, the distributions of $P_T^{\mu_1}$ and ${\:/\!\!\!\! E}$ show two peaks that correspond to the $\mu^{+}\mu^{-}\rightarrow \nu_{\ell}\overline{\nu_{\ell}}\mu^{+}\mu^{-}$ process. The right peak of ${\:/\!\!\!\! E}$ distribution indicates that most of the energy is carried away by the neutrino pair, leaving minimal energy for the two muons, while the left peak indicates that each of the two muons and two neutrinos carries almost an equal share of the energy. For the $\mu^{+}\mu^{-}\rightarrow t\overline{t}\rightarrow (b\mu^+\nu_{\mu})(\overline{b}\mu^-\overline{\nu_{\mu}})$ process, the ${\:/\!\!\!\! E}$ distribution peak is around $2600$ GeV, indicating that the two neutrinos take away more energy. As the signal ${\:/\!\!\!\! E}$ distribution peak is around $2000$ GeV, we applied ${\:/\!\!\!\! E} < 2800$ GeV to reduce these two background events. Moreover, to suppress the $b$ jet background from the $\mu^{+}\mu^{-}\rightarrow t\overline{t}\rightarrow (b\mu^+\nu_{\mu})(\overline{b}\mu^-\overline{\nu_{\mu}})$ process, we vetoed events with $N(b)\ge1$ GeV and $ P_T^b > 25 $ GeV. We also applied the ratio ${\:/\!\!\!\! E}/M_{\mu_1\mu_2}$ as a complementary selection for the $\mu$ALP mass window, setting ${\:/\!\!\!\! E}/M_{\mu_1\mu_2}> 32$. This selection was based on the observation that the position of the average $M_{\mu_1\mu_2}$ distribution of the $\mu^{+}\mu^{-}\rightarrow \nu_{\ell}\overline{\nu_{\ell}}\mu^{+}\mu^{-}$ process is larger than that of the signal, and the ${\:/\!\!\!\! E}$ distribution from this background is relatively small in the range $ 1500 < {\:/\!\!\!\! E} < 2800$ GeV. The $\mu$ALP mass window selection effectively reduced these two backgrounds while keeping most of the signal events. By applying the cuts of $\Delta\phi_{\mu_{1,2}, {\:/\!\!\!\! E}}$ to reduce some events from $\mu^{+}\mu^{-}\rightarrow \nu_{\ell}\overline{\nu_{\ell}}\mu^{+}\mu^{-}$, we observed that two isolated muons were well-separated from ${\:/\!\!\!\! E}$. Especially, the distribution of $\Delta\phi_{\mu_2, {\:/\!\!\!\! E}}$ is not so large in both $\mu^{+}\mu^{-}\rightarrow \nu_{\ell}\overline{\nu_{\ell}}\mu^{+}\mu^{-}$ and $\mu^{+}\mu^{-}\rightarrow t\overline{t}\rightarrow (b\mu^+\nu_{\mu})(\overline{b}\mu^-\overline{\nu_{\mu}})$ compared to the signal. Finally, using a benchmark integrated luminosity ${\cal L} = 120$ fb$^{-1}$ of a muon collider, we defined the signal significance $Z$~\cite{Cowan:2010js} as
\begin{equation}
Z = \sqrt{2\cdot\left( (N_s + N_b)\cdot ln(1+N_s/N_b)-N_s\right)},
\end{equation} 
where $N_s$ and $N_b$ are the relevant signal and background event numbers. 
Here the systematic uncertainties are not taken into account in our simple analysis since the muon collider is still a future collider.
After all of these event selections in Table.~\ref{tab:CF_uu_1}, we find the signal significance can reach $Z=38$ for our benchmark point of ${\cal L} = 120$ fb$^{-1}$ which means $c^A_{\mu} / \Lambda < 0.1$ TeV$^{-1}$ is still detectable in the future.

In the above analysis, we consider the prompt $\mu$ALP decay with the lab frame decay length, $\gamma\beta c\tau_a < 1$ mm as a criterion at a muon collider. Here, $\gamma$ is the Lorentz factor, $\beta$ is the $\mu$ALP velocity, and $\tau_a$ is the proper decay time of $\mu$ALP. However, as we can expect, the $\mu$ALP lab frame decay length becomes longer when $m_a$, $c^A_{\mu}/ \Lambda$ are small, and $\beta$ is large. In this situation, $\mu$ALPs become long-lived particles (LLPs). We take two benchmark points, $m_a =5, 20$ GeV with $c^A_{\mu} / \Lambda = 0.1$ TeV$^{-1}$, to display the $\gamma\beta c\tau_a$ distribution from $\mu^+\mu^-\rightarrow\nu_{\mu}a\overline{\nu_{\mu}}$ at the muon collider in the right panel of Fig.~\ref{fig:L}. We will discuss the situation of $\mu$ALPs as the LLPs later in Sec.~\ref{sec:Summary}.

\begin{table}[ht!]
	\begin{center}\begin{tabular}{|c|c|c|c|}\hline cut flow in $\sigma$ [fb] & ~signal~ & ~$\nu_{\ell}\overline{\nu_{\ell}}c\overline{c}$~ 
			
			& ~$\nu_{\ell}\overline{\nu_{\ell}}b\overline{b}$~ \\ 
			\hline Generator & $2.73$ & $208.20$ & $633.60$ \\
            \hline $\gamma\beta c\tau_a < 1$ mm & $0.52$ & $-$ & $-$ \\
			\hline cut-(1) & $0.50$ & $4.86\times 10^{-3}$ & $0.17$ \\
			\hline cut-(2) & $0.50$ & $1.39\times 10^{-3}$ & $2.41\times 10^{-2}$ \\
			\hline cut-(3) & $0.47$ & $0$ & $6.31\times 10^{-3}$ \\
            \hline cut-(4) & $0.47$ & $0$ & $1.27\times 10^{-3}$ \\
			\hline cut-(5) & $0.42$ & $0$ & $0$ \\ 
			\hline \end{tabular} \caption{Similar to Table.~\ref{tab:CF_uu_1}, but for the signal benchmark point $m_a = 5$ GeV and $c^A_{\mu} / \Lambda = 0.1$ TeV$^{-1}$ as well as the signature of a $J_{\mu}$ candidate plus ${\:/\!\!\!\! E}$.}
		\label{tab:CF_uu_2}
	\end{center}
\end{table}

For the second signal signature, possible SM backgrounds come from $\nu_{\ell}\overline{\nu_{\ell}}b\overline{b}$ and $\nu_{\ell}\overline{\nu_{\ell}}c\overline{c}$ where the heavy flavor mesons produced from $c$, $b$ jets can decay to a collimated muon pair and mimic the $J_{\mu}$ from the signal. The pre-selection cuts ($ P_T^{\mu} > 5 $ GeV, $\lvert\eta_{\mu}\rvert < 2.5$) at parton-level have been used for signal and background processes.  
We take the signal benchmark point as $m_a = 5$ GeV and $c^A_{\mu}/ \Lambda = 0.1$ TeV$^{-1}$. The $C/A$ jet clustering algorithm for $J_{\mu}$ with a cone size $R = 0.1$ which corresponds to the muon isolation criterion at the muon collider is applied. We set up event selections to identify the signal signature and suppress the background events below :  
\begin{itemize}
	\item (1) $N(\mu)\ge 2$ with $ P_T^{\mu_{1,2}} > 5 $ GeV, $\lvert\eta_{\mu_{1,2}}\rvert < 2.5$,
	\item (2) $N(J_{\mu}) = 1$ and $ P_T^{J_{\mu}} > 20 $ GeV, $\lvert\eta_{J_{\mu}}\rvert < 2$ ,
	\item (3) $ 1500 < {\:/\!\!\!\! E} < 2800$ GeV and $\lvert\eta_{{\:/\!\!\!\! E}}\rvert < 1.4$,  
    \item (4) Veto $N(b)\ge1$ GeV with $ P_T^{b} > 25 $ GeV, 
	\item (5) $\lvert M_{J_{\mu}}-m_a\rvert < 2$ GeV. 
\end{itemize} 
The cut-flow table including signal and backgrounds for each event selection is listed in Table.~\ref{tab:CF_uu_2} and some kinematic distributions are shown in Fig.~\ref{fig:10} of Appendix~\ref{app:rec}.

For the $\mu$ALP prompt decay, we first set $\gamma\beta c\tau_a < 1$ mm as a criterion. Then, two muons with $P_T^{\mu} > 5 $ GeV and $\lvert\eta_{\mu}\rvert < 2.5$ are required to be detectable in the muon spectrometer. We consider a $J_{\mu}$ candidate with $P_T^{J_{\mu}} > 20 $ GeV and ${\:/\!\!\!\! E} > 1500$ GeV as the trigger, which are mainly distributed in the central region. The $J_{\mu}$ in signal events comes from energetic $\mu$ALPs, whereas in background events, it comes from the decay of heavy flavor mesons. As shown in Fig.~\ref{fig:10}, the $P_T^{J_{\mu}}$ of the signal is much larger than that of those backgrounds, and most of the background events have been largely reduced after the cut-(2). We further require the selection ${\:/\!\!\!\! E} < 2800$ GeV, which retains most signal events while removing significant parts of background events, particularly the events from $\mu^{+}\mu^{-}\rightarrow\nu_{\ell}\overline{\nu_{\ell}}c\overline{c}$ have been entirely removed. To suppress $\mu^{+}\mu^{-}\rightarrow\nu_{\ell}\overline{\nu_{\ell}}b\overline{b}$, we veto $N(b)\ge1$ GeV with $P_T^{b} > 25 $ GeV. We also require the jet mass of $J_{\mu}$ to satisfy the $\mu$ALP mass window selection, which can entirely remove events from $\mu^{+}\mu^{-}\rightarrow\nu_{\ell}\overline{\nu_{\ell}}b\overline{b}$. After all event selections in Table.~\ref{tab:CF_uu_2}, we can take this signal benchmark point as background-free. The distribution of the peak of $M_{J_{\mu}}$ is broader than that of $M_{\mu_1\mu_2}$ because two muons within a $J_{\mu}$ cannot pass the muon isolation criteria. The selection of the jet clustering method, in conjunction with the choice of cone size $R=0.1$, can affect the four-momentum reconstruction of the $J_{\mu}$. In some cases, one of the muons is outside the jet cone and cannot be reconstructed, leading to distortions in $M_{J_{\mu}}$ compared to $M_{\mu_1\mu_2}$. With ${\cal L} = 120$ fb$^{-1}$, there are 50 signal events left for this benchmark point after all event selections.

\subsection{Exploring $\mu^{+}\mu^{-}\rightarrow \gamma a$ and $\mu^{+}\mu^{-}\rightarrow \mu^{+}\mu^{-} a$ in the \textbf{EWP}  scenario} 
\label{sec:auu}
In the \textbf{EWP} scenario, we employ a different approach to search for $\mu$ALPs compared to the \textbf{EWV} scenario. As explained towards the end of Sec.~\ref{sec:Xsec}, $\mu^+\mu^-\rightarrow\nu_{\mu}a\overline{\nu_{\mu}}$ process in the \textbf{EWP} scenario lacks energy-enhancement behavior, leading to a smaller production cross section. For this reason, we have opted to focus on the following two $\mu$ALP production channels, $\mu^{+}\mu^{-}\rightarrow \gamma a$ and $\mu^{+}\mu^{-}\rightarrow \mu^{+}\mu^{-} a$, which have larger production cross sections, for the signal-to-background analyses in order to obtain stronger future bounds.
We analyzed the process $\mu^{+}\mu^{-}\rightarrow \gamma a$ ($a\rightarrow\mu^{+}\mu^{-}$) using the same method as Sec.~\ref{sec:uuvv}. The details and results are presented below. 
\begin{table}[ht!]
	\begin{center}\begin{tabular}{|c|c|c|}\hline cut flow in $\sigma$ [fb] & ~signal~ & ~$\mu^{+}\mu^{-}\rightarrow \gamma\mu^{+}\mu^{-}$~  \\ 
			\hline Generator & $6.84\times 10^{-2}$ & $179.80$  \\
			\hline cut-(1) & $2.70\times 10^{-2}$ & $4.72$  \\
			\hline cut-(2) & $2.69\times 10^{-2}$ & $0.98$  \\
			\hline cut-(3) & $2.48\times 10^{-2}$ & $0.56$  \\
			\hline cut-(4) & $2.03\times 10^{-2}$ & $2.70\times 10^{-2}$  \\ 
			\hline cut-(5) & $1.68\times 10^{-2}$ & $3.06\times 10^{-3}$  \\
			\hline \end{tabular} \caption{The cut-flow table for $\mu^{+}\mu^{-}\rightarrow \gamma a$ and the relevant SM background with the signature of two isolated muons plus a photon. The benchmark point $m_a = 50$ GeV with $c^A_{\mu} / \Lambda = 10$ TeV$^{-1}$  for the signal is chosen.}
		\label{tab:CF_auu_1.1}
	\end{center}
\end{table}
The signal signatures are first classified into two categories : (1) two isolated muons plus a photon ($\gamma$) for $m_a\gtrsim 15$ GeV, and (2) a $J_{\mu}$ plus a $\gamma$ for $m_a \lesssim 15$ GeV. To investigate the first signal signature, we consider the relevant SM background : $\mu^{+}\mu^{-}\rightarrow \gamma\mu^{+}\mu^{-}$ and choose the benchmark point $m_a = 50$ GeV with $c^A_{\mu} / \Lambda = 10$ TeV$^{-1}$ to display the signal features. The following event selections to identify the signal signature and suppress background events are required :
\begin{itemize}
	\item (1) $N(\mu)\ge 2$ with $ P_T^{\mu_1} > 100 $ GeV,  $10 < P_T^{\mu_2} < 500$ GeV, $\lvert\eta_{\mu_{1,2}}\rvert < 1.5$, 
	\item (2) $ E_{\gamma} > 1450$ GeV and $\lvert\eta_{\gamma}\rvert < 1.6$,
    \item (3) $2.9 < \Delta\phi_{\mu_1, \gamma} < 3.4$ and $2.9 < \Delta\phi_{\mu_2,\gamma} < 3.3$, 
	\item (4) $E_{\gamma}/M_{\mu_1\mu_2} > 29$, 
	\item (5) $\lvert M_{\mu_1\mu_2}-m_a\rvert < 2.0$ GeV.
\end{itemize}
where $E_{\gamma}$ is the energy of photon, $\Delta\phi_{\mu_i,\gamma}$ is the azimuthal angle between the i-th muon and the photon. The cut-flow table detailing the signal and background for each event selection is presented in Table.~\ref{tab:CF_auu_1.1}, with some relevant kinematic distributions shown in Fig.~\ref{fig:auu} of Appendix~\ref{app:rec}.    
First, we found that two isolated muons in the signal is predominantly located in the relatively low transverse momentum regions (as shown in Fig.~\ref{fig:auu}), whereas the number of signal events for $E_{\gamma}$ is mainly located in the relatively higher energy regions. This is because two muons in the signal are produced from the decay of the $\mu$ALP, which are secondary particles. By contrast, two muons from the background events mainly come from the initial particles. Similarly, the photon in the signal is produced from the initial muons and therefore becomes more energetic. On the other hand, for the background process, the energy of photons is more divided by the $P_{z}$ of muons, so the leading photon energy is smaller than that of the signal process as shown in Fig.~\ref{fig:auu}. To select candidate events, we applied the following trigger criteria: $P_T^{\mu_1} > 100$ GeV, $P_T^{\mu_2} > 10$ GeV, and $E_{\gamma} > 1450 $ GeV. To reduce background events, we apply the cut of $\Delta\phi_{\mu_{1,2},\gamma}$, as we observed that two isolated muons are well-separated from the photon. Additionally, we implemented a complementary selection based on the ratio $E_{\gamma}/M_{\mu_1\mu_2}$ to further reduce the contribution from $\mu^+\mu^-\rightarrow\gamma\mu^+\mu^-$. Specifically, we set the ratio $E_{\gamma}/M_{\mu_1\mu_2} > 29$ which effectively reduced background events while retaining most of the signal events.  Since the average position of the $M_{\mu_1\mu_2}$ distribution for the $\mu^{+}\mu^{-}\rightarrow \gamma\mu^{+}\mu^{-}$ process is considerably wider than that of the signal, the $\mu$ALP mass window could further reduce the number of background events. Finally, we consider a benchmark integrated luminosity of a muon collider with ${\cal L} = 1000$ fb$^{-1}$ to our analysis. After all of these event selections in Table.~\ref{tab:CF_auu_1.1}, we find the signal significance can reach $Z=6.379$.

\begin{table}
\begin{center}\begin{tabular}{|c|c|c|c|}\hline cut flow in $\sigma$ [fb] & ~signal~ & ~$\gamma c\overline{c}$~ 
			
			& ~$\gamma b\overline{b}$~ \\ 
			\hline Generator & $8.03\times 10^{-2}$ & $5.96$ & $8.48$ \\
			\hline cut-(1) & $7.78\times 10^{-2}$ & $4.70\times 10^{-3}$ & $1.78\times 10^{-2}$ \\
			\hline cut-(2) & $6.36\times 10^{-2}$ & $7.63\times 10^{-4}$ & $5.49\times 10^{-4}$ \\
			\hline cut-(3) & $5.29\times 10^{-2}$ & $1.19\times 10^{-5}$ & $1.50\times 10^{-4}$ \\
            \hline cut-(4) & $4.43\times 10^{-2}$ & $0$ & $1.33\times 10^{-4}$ \\ 
            \hline cut-(5) & $3.61\times 10^{-2}$ & $0$ & $4.99\times 10^{-5}$ \\
			\hline \end{tabular} \caption{Similar to Table.~\ref{tab:CF_auu_1.1}, but for the signal benchmark point $m_a = 5$ GeV and $c^A_{\mu} / \Lambda = 10$ TeV$^{-1}$ as well as the signature of a $J_{\mu}$ candidate plus a photon.}
		\label{tab:CF_uu_5}
	\end{center}
\end{table}

For the second signal signature, possible SM backgrounds come from $\gamma b\overline{b}$ and $\gamma c\overline{c}$ where the heavy flavor mesons generated from $c$ and $b$ jets can decay into a collimated muon pair and mimic the $J_{\mu}$ from the signal. The pre-selection cuts ($ P_T^{\mu} > 5 $ GeV, $\lvert\eta_{\mu}\rvert < 2.5$) at parton-level have been used for signal and background processes.  We still choose the signal benchmark point as $m_a = 5$ GeV and $c^A_{\mu}/ \Lambda = 10$ TeV$^{-1}$. Applying the $C/A$ jet clustering algorithm for $J_{\mu}$ with a cone size $R = 0.1$, we implement event selection criteria to isolate the signal and suppress background events as specified below :  
\begin{itemize}
	\item (1) $N(\mu)\ge 2$ with $ P_T^{\mu_{1,2}} > 5 $ GeV, $\lvert\eta_{\mu_{1,2}}\rvert < 2.5$,
	\item (2) $N(J_{\mu}) = 1$ and $ P_T^{J_{\mu}} > 750 $ GeV, $\lvert\eta_{J_{\mu}}\rvert < 1.5$ ,
	\item (3) $ E_{\gamma} > 1400 $ GeV and $\lvert\eta_{\gamma}\rvert < 1.0$,  
    \item (4) $\lvert M_{J_\mu}-m_a\rvert < 3.0$ GeV, 
    \item (5) $100 < E_{\gamma}/M_{J_\mu} < 400$.
\end{itemize} 
The reasons for implementing these event selection criteria are similar to that described earlier, and will not be repeated here again. A cut-flow table detailing the signal and backgrounds for each event selection is presented in Table.~\ref{tab:CF_uu_5} with some relevant kinematic distributions shown in Fig.~\ref{fig:j10} of Appendix~\ref{app:rec}. Employing all the event selections listed in Table.~\ref{tab:CF_uu_2}, we find that the signal significance can achieve a value of $Z=20.10$ with ${\cal L} = 1000$ fb$^{-1}$.

\begin{table}[ht!]
	\begin{center}\begin{tabular}{|c|c|c|}\hline cut flow in $\sigma$ [fb] & ~signal~ & ~$\mu^{+}\mu^{-}\rightarrow \mu^{+}\mu^{-}\mu^{+}\mu^{-}$~  \\ 
			\hline Generator & $1.06\times 10^{-3}$ & $2.82$  \\
            \hline cut-(1) & $1.03\times 10^{-3}$ & $1.26$  \\
			\hline cut-(2) & $6.53\times 10^{-4}$ & $3.26\times 10^{-1}$  \\
			\hline cut-(3) & $4.74\times 10^{-4}$ & $1.59\times 10^{-1}$  \\
            \hline cut-(4) & $4.35\times 10^{-4}$ & $5.80\times 10^{-2}$ \\
            \hline cut-(5) & $3.43\times 10^{-4}$ & $3.67\times 10^{-3}$ \\
			\hline \end{tabular} \caption{Similar to Table.~\ref{tab:CF_auu_1.1}, but for the $\mu^{+}\mu^{-}\rightarrow \mu^{+}\mu^{-} a$ channal and relevant SM backgrounds with the signature of four isolated muons.}
		\label{tab:CF_zuu_1}
	\end{center}
\end{table} 

In the subsequent section, we analyze the process $\mu^{+}\mu^{-}\rightarrow \mu^{+}\mu^{-} a$ (include $\mu^{+}\mu^{-}\rightarrow Z a\rightarrow (\mu^+\mu^-) a$) using the same method as Sec.~\ref{sec:uuvv}.
The signal signatures are classified into two categories : (1) four isolated muons for $m_a \gtrsim 15$ GeV, and (2) a $J_{\mu}$ plus two isolated muons for $m_a \lesssim 15$ GeV. To investigate the first signal signature, we consider the relevant SM background: $\mu^{+}\mu^{-}\rightarrow \mu^{+}\mu^{-}\mu^{+}\mu^{-}$ and choose the same signal benchmark point $m_a = $ 50 GeV with $c_{\mu}^A / \Lambda = 10 $ TeV$^{-1}$ to display the signal features. The following event selections to identify the signal signature and suppress background events are required :
\begin{itemize}
	\item (1) $N(\mu)\ge 4$ with $ P_T^{\mu_{1,2,3,4}} > 5 $ GeV, $\lvert\eta_{\mu_{1,2,3,4}}\rvert < 2.5$,
    \item (2) $ P_T^{\mu_1} > 200 $ GeV, $ P_T^{\mu_{2,3}} > 100 $ GeV, $\lvert\eta_{\mu_{1}}\rvert < 2.0$ and $\lvert\eta_{\mu_{4}}\rvert < 1.5$,
	\item (3) $\Delta\phi_{\mu_2,\mu_4} > 0.5$ and $\Delta\phi_{\mu_3,\mu_4} > 1$,
    \item (4) $P_T^{\mu_4}/M_{\mu_1 \mu_4} > 0.05$,
    \item (5) $\lvert M_{\mu_1\mu_4}-m_a\rvert < 5.0$ GeV.
\end{itemize} 
The cut-flow table including the signal and the background for each event selection is listed in Table.~\ref{tab:CF_zuu_1} and some kinematic distributions are shown in Fig.~\ref{fig:zuu} of Appendix~\ref{app:rec}.

Four isolated muons with $P^{\mu}_T > 5$ GeV and $\lvert\eta_{\mu}\rvert < 2.5$ are applied as a trigger criteria. 
We observed that $P^{\mu_1}_T$, $P^{\mu_2}_T$ and $P^{\mu_3}_T$ of the signal are more energetic than that of the background. On the other hand, $\mu_1$ and $\mu_4$ from the signal are distributed in the central regions relative to that from the background which $\mu_1$ and $\mu_4$ are mainly generated by the initial muons with the forward and backward directions. Therefore, we choose the cut-(2) in the above to select candidate events. 
We have checked all combinations of four muons in the final state to reconstruct a pair of muons which comes from the $\mu$ALP decay and found the pair of $\mu_1$ and $\mu_4$ is most likely to reconstruct the mass of $\mu$ALP. To reduce background events, we apply the event selections based on $\Delta\phi_{\mu_{2,4}}$ and $\Delta\phi_{\mu_{3,4}}$. We observe that the muons produced by the decay of $\mu$ALPs in the signal are well-separated from $\mu_2$ and $\mu_3$. Furthermore, we incorporate an additional selection criterion involving the ratio $P_T^{\mu_4}/M_{\mu_1 \mu_4}$ to further suppress the contribution from $\mu^{+}\mu^{-}\rightarrow \mu^{+}\mu^{-}\mu^{+}\mu^{-}$. Specifically, we set the ratio $P_T^{\mu_4}/M_{\mu_1 \mu_4} > 0.05$, which effectively reduced background events while retaining the majority of signal events. Since the $M_{\mu_{1}\mu_{4}}$ distribution for the $\mu^{+}\mu^{-}\rightarrow \mu^{+}\mu^{-}\mu^{+}\mu^{-}$ process is almost concentrated in the region larger than $85$ GeV, the $\mu$ALP mass window could further reduce the number of background events. Finally, after all of these event selections in Table.~\ref{tab:CF_zuu_1}, we find the signal significance can reach $Z=10.44$ for ${\cal L} = 1000$ fb$^{-1}$ in our analysis.

\begin{table}
\begin{center}\begin{tabular}{|c|c|c|c|}\hline cut flow in $\sigma$ [fb] & ~signal~ & ~$ \mu^{+}\mu^{-}c \overline{c}$~ & ~$\mu^{+}\mu^{-}b \overline{b}$~ \\ 
			\hline Generator & $1.26\times 10^{-3}$ & $52.94$ & $90.18$ \\
			\hline cut-(1) & $1.13\times 10^{-3}$ & $1.98\times 10^{-3}$ & $7.01\times 10^{-2}$ \\
            \hline cut-(2) & $7.19\times 10^{-4}$ & $5.82\times 10^{-4}$ & $2.12\times 10^{-2}$ \\
			\hline cut-(3) & $6.63\times 10^{-4}$ & $2.65\times 10^{-4}$ & $1.52\times 10^{-2}$ \\
            \hline cut-(4) & $6.61\times 10^{-4}$ & $1.06\times 10^{-4}$ & $1.25\times 10^{-2}$ \\
            \hline cut-(5) & $6.03\times 10^{-4}$ & $0$ & $8.21\times 10^{-3}$ \\
            \hline cut-(6) & $5.39\times 10^{-4}$ & $0$ & $5.41\times 10^{-4}$ \\
			\hline \end{tabular} 
   \caption{Similar to Table.~\ref{tab:CF_uu_5}, but for the $\mu^+\mu^-\rightarrow\mu^+\mu^- a$ channel and relevant SM backgrounds with the signature of a $J_{\mu}$ candidate plus two isolated muons.}
		\label{tab:CF_zuu_2}
	\end{center}
\end{table}

For the second signal signature, possible SM backgrounds come from $\mu^{+}\mu^{-}c \overline{c}$ and $\mu^{+}\mu^{-}b \overline{b}$. The pre-selection cuts ($P_T^{\mu} > 5$ GeV, $\lvert\eta_{\mu_{}}\rvert < 2.5$) at parton-level have been used for signal and background process. We still choose the signal benchmark point as $m_a = 5$ GeV and $c^A_{\mu} / \Lambda = 10$ TeV$^{-1}$. Applying the same $C/A$ jet clustering algorithm for a $J_{\mu}$ with a cone size $R = 0.1$, we set up event selections to pick up the signal and suppress background events as specified below :
\begin{itemize}
	\item (1) $N(\mu)\ge 4$ with  $ P_T^{\mu_{1,2,3,4}} > 5 $ GeV, $\lvert\eta_{\mu_{1,2,3,4}}\rvert < 2.5$, 
    \item (2) $ P_T^{\mu_{2,3}} > 100 $ GeV, $\lvert\eta_{\mu_{2,3}}\rvert < 2.0$, 
	\item (3) $N(J_{\mu}) = 1$ with $300 < P_T^{J_{\mu}} < 1400 $ GeV, $\lvert\eta_{J_\mu}\rvert < 1.6$,
	\item (4) $\Delta\phi_{J_{\mu},\mu_2} > 4.5$ and $1.0 < \Delta\phi_{\mu_{2},\mu_3} < 5.0$, 
    \item (5) $P_{T}^{\mu_2}/M_{J_{\mu}} < 500$,
    \item (6) $\lvert M_{J_{\mu}}-m_{a}\rvert < 4$ GeV. 
\end{itemize} 
The event selection criteria have been implemented for reasons similar to those previously described and will not be reiterated here. The cut-flow table, which includes both signal and background events for each selection, is presented in Table.~\ref{tab:CF_zuu_2}, while the corresponding kinematic distributions can be found in Fig.~\ref{fig:zuu_jet_all} of Appendix~\ref{app:rec}. Upon applying all of the event selections outlined in Table.~\ref{tab:CF_zuu_2}, we observe a significant signal significance of $Z=19.86$ for our benchmark point with ${\cal L} = 1000$ fb$^{-1}$.

\subsection{Main results and existing bounds} 
\label{sec:Summary}

\begin{figure*}[ht!]
\includegraphics[width=1.0\textwidth]{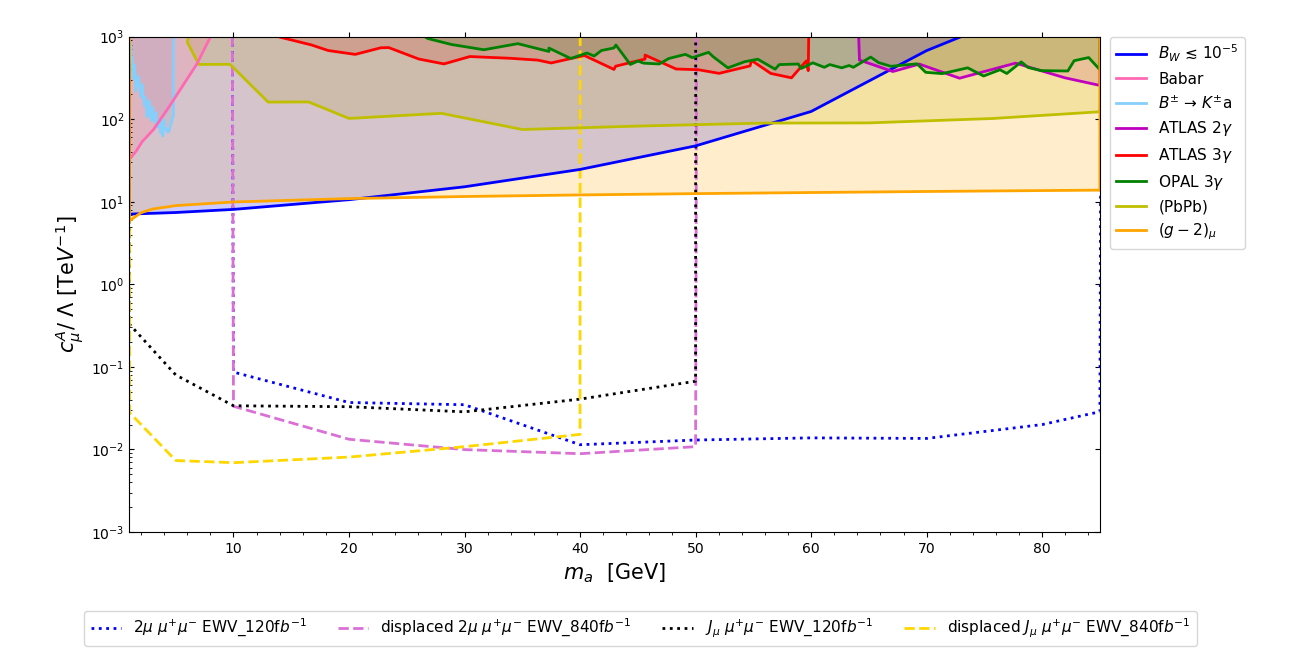}
\caption{The future bounds on $c^A_\mu/\Lambda$ of GeV-scale $\mu$ALPs in the \textbf{EWV} scenario from the muon collider with ${\cal L} = 120$ fb$^{-1}$ and  ${\cal L} = 840$ fb$^{-1}$ within $95\%$ CL or 10 survival events for background-free cases (dotted lines for the $\mu$ALP prompt decay and dashed lines for the $\mu$ALP as a LLP) as well as existing bounds (bulk regions). Here we lable "$2\mu$" and "$J_{\mu}$" to identify two kinds of signatures at a muon collider. ${\cal B}_W\lesssim 10^{-5}$ represents ${\cal B} (W^{\pm}\rightarrow\mu^{\pm}\nu_{\mu}a) < 10^{-5}$~\cite{Altmannshofer:2022izm} (blue bulk). For light $\mu$ALPs, BaBar~\cite{BaBar:2016sci} (hotpink bulk), $B^{\pm}\rightarrow K^{\pm}a\rightarrow K^{\pm}(\gamma\gamma)$~\cite{BaBar:2021ich} (lightskyblue bulk) are considered. Some other collider bounds are in order : ATLAS $2\gamma$~\cite{ATLAS:2014jdv,Jaeckel:2015jla,Knapen:2016moh} (magenta bulk), ATLAS $3\gamma$~\cite{ATLAS:2015rsn,Knapen:2016moh} (red bulk), OPAL $3\gamma$~\cite{OPAL:2002vhf,Knapen:2016moh} (green bulk), ATLAS/CMS (PbPb)~\cite{dEnterria:2021ljz} (yellow bulk). Finally, the bound from $(g-2)_{\mu}$~\cite{Ganguly:2022imo} is labeled as orange bulk.}
\label{fig:summary}
\end{figure*} 

\begin{figure*}[ht!]
\includegraphics[width=0.95\textwidth]{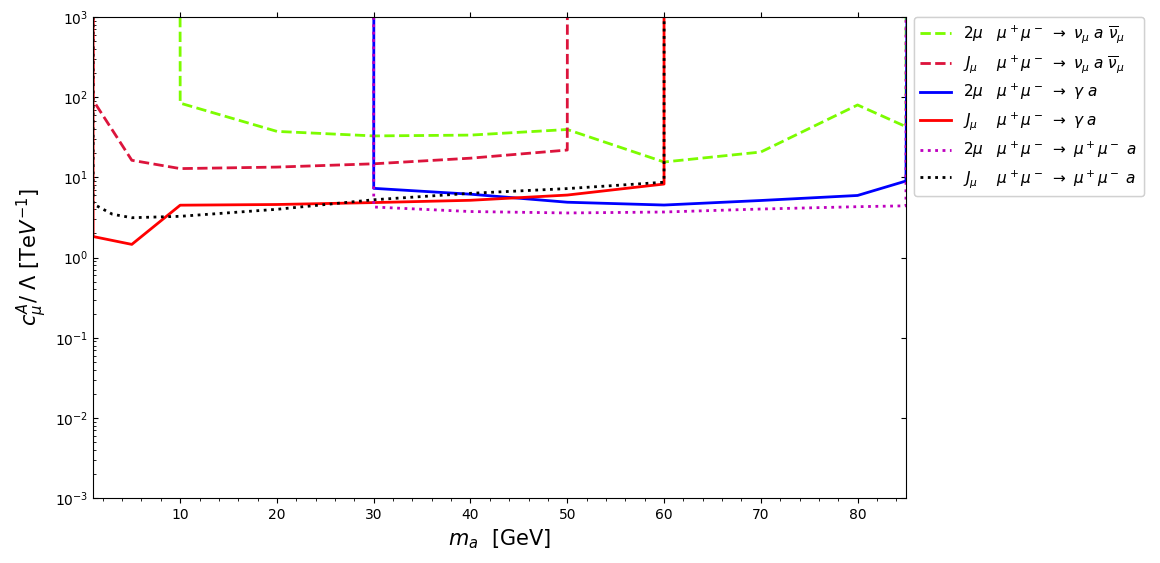}
\caption{The future bounds on $c^A_\mu/\Lambda$ of GeV-scale $\mu$ALPs from the muon collider in the \textbf{EWP} scenario with ${\cal L} = 1000$ fb$^{-1}$  within $95\%$ CL or 10 survival events for background-free cases. Only the bounds from the $\mu$ALP prompt decay are considered (dashed lines for $\mu^+\mu^-\rightarrow\nu_{\mu}a\overline{\nu_{\mu}}$, solid lines for $\mu^{+}\mu^{-}\rightarrow \gamma a$ as well as dotted lines for $\mu^{+}\mu^{-}\rightarrow \mu^{+}\mu^{-} a$). The labels of "$2\mu$" and "$J_{\mu}$" are the same as Fig.~\ref{fig:summary}.}
\label{fig:EWP}
\end{figure*} 

The study of signal benchmark points is extended to a wide range of $m_a$ by employing the search strategies outlined in Sec.~\ref{sec:uuvv} and Sec.~\ref{sec:auu}, resulting in the identification of possible future bounds within $95\%$ confidence level (CL) ($Z=1.96$). To conservatively demonstrate the signal significance of the case without the survival background event after all event selections or the case of background-free assumption, a minimum of $10$ signal events is required to be present, and only signal efficiency larger than $10\%$ is considered in the analysis. Our study is restricted to $1$ GeV $\le m_a \lesssim M_W$ for $\mu$ALPs below the electroweak scale. We first summarize our results in Fig.~\ref{fig:summary} for searching the $\mu^{+}\mu^{-}\rightarrow\nu_{\mu} a \overline{\nu}_{\mu}$ channel in the \textbf{EWV} scenario. The dotted lines are used for the case of prompt $\mu$ALPs decay ($\gamma\beta c\tau_a < 1$ mm) at a muon collider with ${\cal L} = 120$ fb$^{-1}$. Note the lower bound of $m_a$ comes from technical issues of $J_{\mu}$ analysis. When $m_a < 1$ GeV, the $\mu$ALP mass window selection is no longer powerful to distinguish the signal from backgrounds. On the other hand, since the event selections for two isolated muons plus ${\:/\!\!\!\! E}$ at the muon collider are sensitive to the values of $m_a$, event selections are dynamically optimized for different $m_a$ to suit each case as shown in Table.~\ref{tab:other}. 
The case of prompt decay of $\mu$ALPs with two isolated muons at the muon collider yields a background-free scenario when $m_a\lesssim 30$ GeV, causing a cusp point at $m_a = 30$ GeV in Fig.~\ref{fig:summary}. Similarly, for the prompt decay of $\mu$ALPs with a $J_{\mu}$ at the muon collider, SM background events can only survive after all event selections when $m_a = 10$ GeV, resulting in a cusp point at $m_a = 10$ GeV in Fig.~\ref{fig:summary}.

For $\mu$ALPs as the LLPs, we first consider the physical size in radius of proposed detectors for muon colliders~\cite{MuonCollider:2022ded}. Some relevant detector parameters for the inner and outer radius of the vertex detector, ECAL, HCAL and muon system are summed up as follows : (1) $3.0 \leq R_{\text{vertex}} \leq 10.4$ cm, (2) $150.0 \leq R_{\text{ECAL}} \leq 170.2$ cm, (3) $174.0 \leq R_{\text{HCAL}} \leq 333.0$ cm, (4) $446.1 \leq R_{\text{muon}} \leq 645.0$ cm. Therefore, we simply consider the $\mu$ALP lab frame decay length within $10^{-3} \leq\gamma\beta c \tau_a\leq 6.4$ m as a detectable LLP with a muon pair displaced vertex and a displaced $J_{\mu}$ signatures at a muon collider. We assume that both a muon pair displaced vertex and a displaced $J_{\mu}$ signatures at muon colliders are background-free after the trigger and $\mu$ALP mass window selection implementation, as described in the previous text. The analysis of a muon pair displaced vertex and a displaced $J_{\mu}$ signatures at the muon collider is carried out using an integrated luminosity of ${\cal L} = 840$ fb$^{-1}$, respectively. The results of the LLP study for searching the $\mu^{+}\mu^{-}\rightarrow\nu_{\mu} a \overline{\nu}_{\mu}$ channel in the \textbf{EWV} scenario are summarized in Fig.~\ref{fig:summary} with the dashed lines. The signal efficiency of the two isolated muon signature decreases when $m_a\lesssim 30$ GeV because these two muons become too close to each other and cannot pass the muon isolation criterion. Similarly, grouping two muons inside a $J_{\mu}$ candidate is challenging for $m_a\gtrsim 10$ GeV at the muon collider. Therefore, the analysis of signatures with two isolated muons and a $J_{\mu}$ complement each other for $\mu$ALP searches in the middle $m_a$ range.

Some existing bounds are also shown in Fig.~\ref{fig:summary} for the comparison. First of all, according to the interaction in Eq.~(\ref{eq:Jint}), there is a new $W$ boson exotic decay channel, $W^{+}\rightarrow\mu^{+}\nu_{\mu}a$.  The precision measurements of $W$ boson width ($\Gamma_W = 2.085\pm 0.042$ GeV~\cite{Workman:2022ynf}) can indirectly test $\mu$ALP with $m_a < M_W$ in the \textbf{EWV} scenario. Here we conservatively require the branching ratio of $W^{+}\rightarrow\mu^{+}\nu_{\mu}a$ to be less than $10^{-5}$~\cite{Altmannshofer:2022izm} and mark it as the blue bulk in Fig.~\ref{fig:summary}. 
For lighter $\mu$ALPs ($m_a\lesssim 5$ GeV), searching for four muons in the final state~\cite{BaBar:2021ich} (hotpink bulk) and $B^{\pm}\rightarrow K^{\pm}a$ (light skyblue bulk) by BaBar experiments can already constrain some parameter space in the upper-left corner. 
For heavier $\mu$ALPs ($m_a > 5$ GeV), the ATLAS $2\gamma$~\cite{ATLAS:2014jdv,Jaeckel:2015jla,Knapen:2016moh} (magenta bulk), ATLAS $3\gamma$~\cite{ATLAS:2015rsn,Knapen:2016moh} (red bulk), OPAL $3\gamma$~\cite{OPAL:2002vhf,Knapen:2016moh} (green bulk), ATLAS/CMS (PbPb)~\cite{dEnterria:2021ljz} (yellow bulk) can already exclude some parameter space with $c^A_{\mu} / \Lambda\gtrsim 10^2$ TeV$^{-1}$. 
 On the other hand, the precision measurements of muon magnetic moment can also provide constraints for $\mu$ALPs. The combined measurement from Fermilab and Brookhaven is reported as $a^{\text{EXP}}_{\mu}=116,592,061(41)\times 10^{-11}$~\cite{Muong-2:2021ojo} and if we consider the lattice calculation for hadronic vacuum polarization (HVP), the SM prediction value change to $a^{\text{SM}}_{\mu}=116,591,954(55)\times 10^{-11}$~\cite{Borsanyi:2020mff}. In this situation, the deviation of $(g-2)_{\mu}$ is reported as $\Delta a_{\mu} = a^{\text{EXP}}_{\mu} -a^{\text{SM}}_{\mu} = 107(69)\times 10^{-11}$ and we consider the $\Delta a_{\mu}$ observation within $2\sigma$ for $\mu$ALPs in this work. The one-loop contributions from light $\mu$ALPs to $(g-2)_{\mu}$ is negative and can be written as\footnote{The one-loop contribution from $aZ\gamma$ interaction and the two-loop contribution from $aW^{+}W^{-}$ as well as the two-loop light-by-light contribution are much suppressed compared with Eq.~(\ref{eq:mug2_1}). Hence, we can safely ignore their effects here.} 
 
\begin{equation}
\Delta a^{\rm 1-loop}_{\mu} = \Delta a^{\mu a\mu}_{\mu} +\Delta a^{\mu a\gamma}_{\mu}\,\,\,, 
\label{eq:mug2_1}
\end{equation}
where the first term comes from the $\mu$-$a$-$\mu$ loop and the second term comes from the $\mu$-$a$-$\gamma$ loop as shown in Ref.~\cite{Ganguly:2022imo} for the following form, 
\begin{align}
\Delta a^{\mu a\mu}_{\mu} = & -\left(\frac{c^A_{\mu}m_\mu}{\Lambda}\right)^2\frac{r}{8\pi^2}
\int_0^1 dx \dfrac{x^3}{1-x+rx^2}\,\,\,, \\
\Delta a^{\mu a\gamma}_{\mu} = & -\frac{\alpha_{\rm em}}{4\pi^3}
\left(\frac{c^A_{\mu}m_\mu}{\Lambda}\right)^2\times \notag \\
& \int_0^1 dx \left[
(1-x)\left(\ln \dfrac{\Lambda^2_{\rm loop}}{\Delta^2}-\dfrac{1}{2}\right) 
-3r\left\{x^2 \ln\left(\frac{rx^2 +(1-x)}{rx^2}\right)\right\}
\right]\,\,.
\label{eq:mug2_2}
\end{align}
Here $r=m^2_{\mu}/m^2_a$, $\Delta^2 = m^2_{\mu}x^2 +m^2_a (1-x)$ and $\Lambda_{\rm loop}$ is the cut-off scale of the loop integration which is taken to be $1$ TeV here. 
The strongest constraint among all the above ones is from $(g-2)_{\mu}$~\cite{Ganguly:2022imo} (orange bulk), with $c^A_{\mu} / \Lambda\gtrsim 10$ TeV$^{-1}$ and extending to a wide range of $m_a$. It is important to note that all the above bounds have been rescaled according to our definition of ALP-muon interactions in Eq.~(\ref{eq:int}) and $\mu$ALP decay branching ratios in Fig.~\ref{fig:lALP_BR}. However, some of the other bounds such as OPAL $2\gamma$~\cite{OPAL:2002vhf,Knapen:2016moh}, Belle II~\cite{Belle-II:2020jti}, and LHCb~\cite{CidVidal:2018blh}, are so weak that we have not included them here. In comparison to these existing bounds, our proposals to search for $\mu$ALPs via $\mu^+\mu^-\rightarrow\nu_{\mu}a\overline{\nu_{\mu}}$ at muon colliders are still attractive. Furthermore, the possible future bounds of $c^A_\mu/\Lambda$ can reach less than $0.01-0.1$ TeV$^{-1}$, which open new doors to explore $m_a$ in the \textbf{EWV} scenario below the electroweak scale. 

In addition, as we have discussed in Sec.~\ref{sec:Xsec}, in the \textbf{EWV} scenario, cross sections are more than six orders of magnitude larger than those in the \textbf{EWP} scenario for $\mu^{+}\mu^{-}\rightarrow\nu_{\mu} a\overline{\nu_{\mu}}$ processes in Fig.~\ref{fig:ALP_Xsec}. Therefore, future bounds from this channel in the \textbf{EWP} scenario are as small as existing bounds. Additionally, almost the entire cross-section comes from $\boldsymbol{aVV'}$ interaction in the \textbf{EWP} scenario at the muon collider. Comparing the \textbf{EWP} scenario with the \textbf{EWV} one, the longitudinal momentum ($P_z$) becomes larger than the transverse momentum ($P_T$) for two isolated muon pair, because the dominant contribution in the signal process is $\mu^+\mu^-\rightarrow Za\rightarrow (\nu\overline{\nu})(\mu^+\mu^-)$ instead of the one from the four-point interaction. When $m_a \gtrsim 30 $ GeV, the total energy will be roughly equally divided into $Z$ and the ALP, resulting in large changes in some kinematic distributions. In order for comparison, we used the same event selections for both \textbf{EWV} and \textbf{EWP} scenarios. Most of the signal efficiencies are below $10\%$ in the \textbf{EWP} scenario because the condition $P_T^{\mu_1} > 200 $ GeV is too stringent in this situation. Meanwhile, the efficiency of the signal is also very sensitive to ${\:/\!\!\!\! E}/M_{\mu_1\mu_2}$. Thus, as we can expect, the distributions of two isolated muons in \textbf{EWP} scenario are distinct from the ones in the \textbf{EWV} scenario. Eventually, the \textbf{EWP} signal efficiency is about $10\%$ to $40\%$ less than that of the \textbf{EWV} one. At the same time, we explore the potential results of searching for $\mu$ALPs in different channels with ${\cal L} = 1000$ fb$^{-1}$. Except for $\mu^+\mu^-\rightarrow\nu_{\mu}a\overline{\nu_{\mu}}$, we also include $\mu^{+}\mu^{-}\rightarrow \gamma a$, and  $\mu^{+}\mu^{-}\rightarrow \mu^{+}\mu^{-} a$ in the \textbf{EWP} scenario. Due to different generation mechanisms among these channels, the coverage range of the interval of $m_a$ may vary. The case of prompt decay of $\mu$ALPs with two isolated muons in the $\mu^{+}\mu^{-}\rightarrow \gamma a$ channel yields a background-free scenario when $m_a\gtrsim 10$ GeV, causing a cusp point at $m_a = 10$ GeV in Fig.~\ref{fig:EWP}. The case of prompt decay of $\mu$ALPs with four isolated muons in the $\mu^{+}\mu^{-}\rightarrow \mu^{+}\mu^{-} a$ channel causes a cusp point at $m_a = 5$ GeV in Fig.~\ref{fig:EWP}. Finally, we find that a photon plus a $\mu$ALP channel shows the best potential for searching for $\mu$ALPs in the \textbf{EWP} scenario. The possible future bounds on $c^A_\mu/\Lambda$ can reach values less than $1-10$ TeV$^{-1}$, which is only slightly greater than existing bounds.

\section{Conclusions}
\label{sec:final}

Axion-like particles (ALPs) are pseudo-Nambu Goldstone bosons that exist beyond the standard model (SM). In the effective field theory framework, ALPs are allowed to have masses ranging from nearly massless to the electroweak scale or higher, and their couplings with SM particles can be investigated independently. Therefore, it is crucial to search for ALPs with various mass ranges and interaction types. This study focuses on exploring the search for the GeV-scale muonphilic ALPs ($\mu$ALPs), a specific type of ALPs that interact predominantly with muons, at a muon collider.

Producing GeV-scale $\mu$ALPs is challenging due to their suppressed production cross sections, which are proportional to the square of the muon mass. Hence, a new proposal is necessary to produce them effectively at high-energy colliders. This study proposes four production channels that can be used to search for $\mu$ALPs at muon colliders : $\mu^+\mu^-\rightarrow\nu_{\mu}a\overline{\nu_{\mu}}$, $\mu^{+}\mu^{-}\rightarrow \gamma a$, $\mu^{+}\mu^{-}\rightarrow Z a$ and $\mu^{+}\mu^{-}\rightarrow \mu^{+}\mu^{-} a$ which rely on a four-point interaction, $W$-$\mu$-$\nu_{\mu}$-$a$, and/or interactions arising from the chiral anomaly that do not depend on the muon mass. It is noteworthy that in the electrowek
violating (\textbf{EWV}) scenario, the cross section of the $\mu^+\mu^-\rightarrow\nu_{\mu}a\overline{\nu_{\mu}}$ process is six to seven orders of magnitude larger than that of other channels, as shown in Table.~\ref{tab:other}, due to the energy enhancement behavior resulting from the $W$-$\mu$-$\nu_{\mu}$-$a$ interaction. However, in the electroweak preserving (\textbf{EWP}) scenario, the four-point interaction disappears, and the $\mu^{+}\mu^{-}\rightarrow \gamma a$ channel has the largest cross section.

In the search for GeV-scale $\mu$ALPs at a muon collider, different search strategies are employed for the \textbf{EWV} and \textbf{EWP} scenarios. 
We focus on the production channels $\mu^+\mu^-\rightarrow\nu_{\mu}a\overline{\nu_{\mu}}$ in the \textbf{EWV} scenario, and $\mu^{+}\mu^{-}\rightarrow \gamma a$ and $\mu^{+}\mu^{-}\rightarrow \mu^{+}\mu^{-} a$ in the \textbf{EWP} scenario. 
On the other hand, the GeV-scale $\mu$ALP mainly decays into a pair of muons. When the light $\mu$ALP is highly boosted and produced at a muon collider, these two muons are too collimated to pass standard muon isolation criteria and form a novel object called a muon-jet, $J_{\mu}$. Therefore, this study explores two types of signatures : (1) two isolated muons plus other parts, and (2) a $J_{\mu}$ plus other parts. These two signature types are complementary in the search for the GeV-scale $\mu$ALP. The signature of $J_{\mu}$ can cover low-mass $\mu$ALP detection range well, and the signature of two isolated muons can cover high-mass $\mu$ALP detection range. After a comprehensive signal-to-background analysis for these two kind of signatures at a muon collider, future bounds for $c^A_\mu/\Lambda$ are shown to be more than three orders of magnitude stronger than existing bounds for $\mu$ALPs with $1$ GeV $\leq m_a\lesssim M_W$ at integrated luminosity of ${\cal L} = 120$ fb$^{-1}$ for the prompt $\mu$ALP decay and ${\cal L} = 840$ fb$^{-1}$ for the $\mu$ALP as a long-lived particle in the \textbf{EWV} scenario, as illustrated in Fig.~\ref{fig:summary}. However, future bounds for $c^A_\mu/\Lambda$ of $\mu$ALPs with $1$ GeV $\leq m_a\lesssim M_W$ are shown to be barely exceed existing bounds in the \textbf{EWP} scenario, even with an integrated luminosity of ${\cal L} = 1000$ fb$^{-1}$, as illustrated in Fig.~\ref{fig:EWP}. 
Overall, this study provides important insights into the potential to explore GeV-scale $\mu$ALPs. Such efforts will motivate experimentalists to pursue $\mu$ALP searches at future muon colliders. 

\appendix
\section{Some kinematic distributions and supplemental information}
\label{app:rec}
\begin{figure*}[ht!]
\centering{\includegraphics[width=0.48\textwidth]{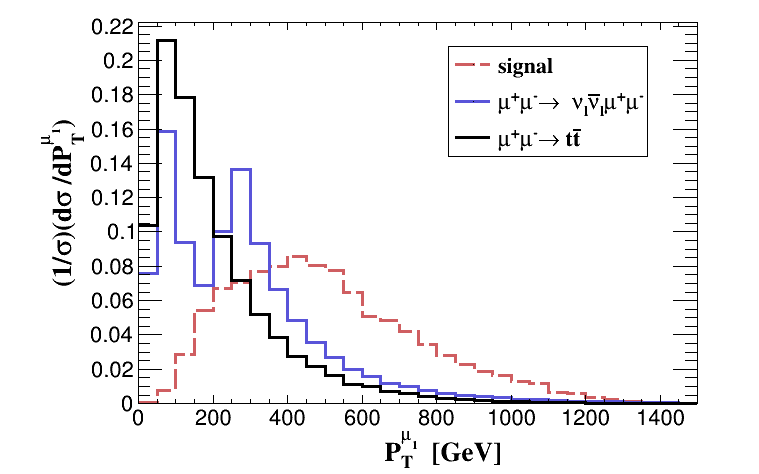}}
\centering{\includegraphics[width=0.48\textwidth]{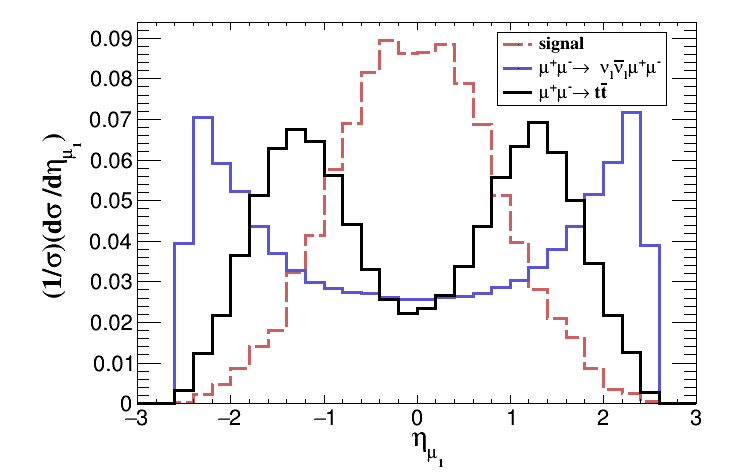}}
\centering{\includegraphics[width=0.48\textwidth]{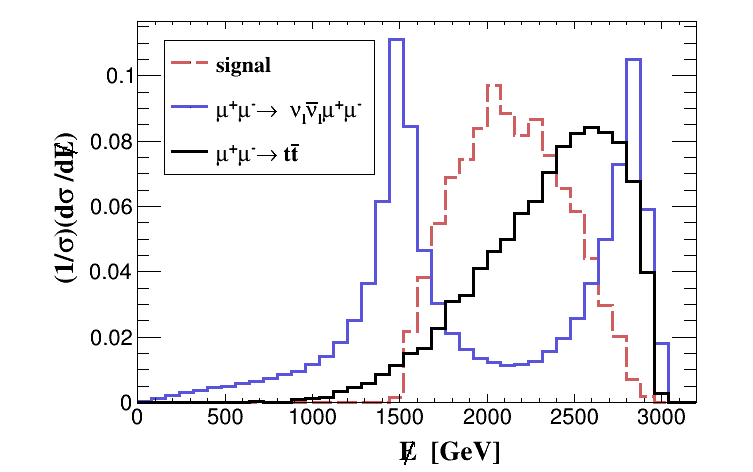}}
\centering{\includegraphics[width=0.48\textwidth]{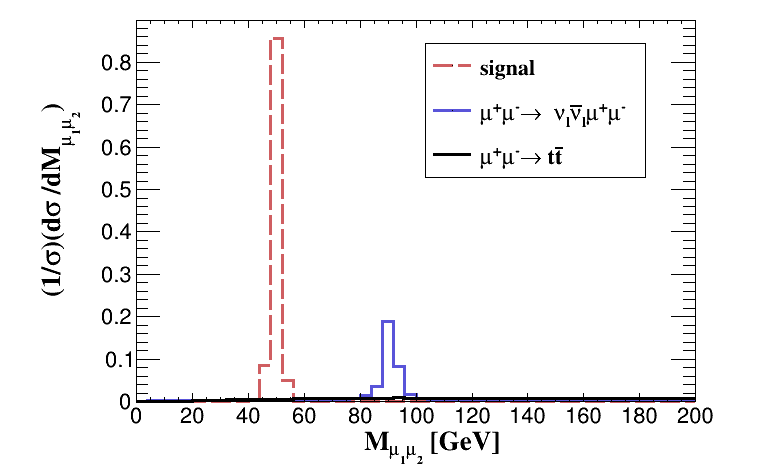}}
\centering{\includegraphics[width=0.48\textwidth]{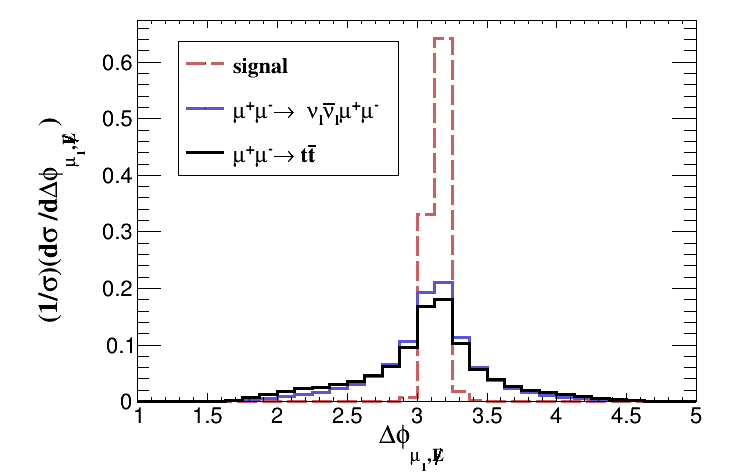}}
\centering{\includegraphics[width=0.48\textwidth]{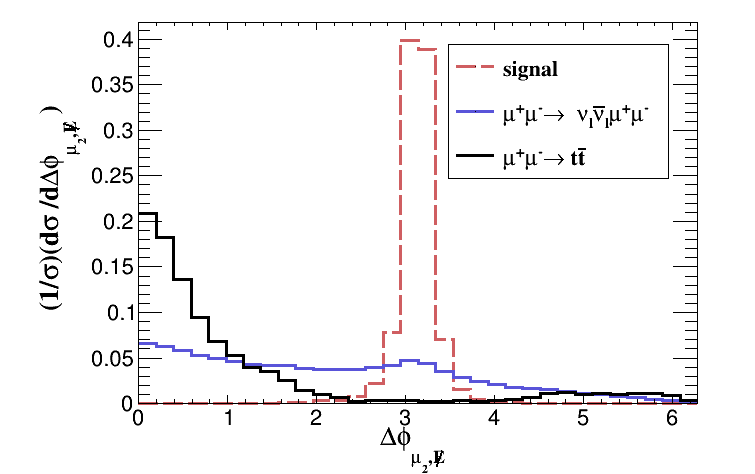}}
\caption{Some signal and background kinematic distributions for the signature of two isolated muons plus ${\:/\!\!\!\! E}$ at a $\mu^+\mu^-$  collider, $P_T^{\mu_1}$, $\eta_{\mu_1}$, ${\:/\!\!\!\! E}$, $M_{\mu_1\mu_2}$, $\Delta\phi_{\mu_1,{\:/\!\!\!\! E}}$ and $\Delta\phi_{\mu_2,{\:/\!\!\!\! E}}$ distributions for $m_a = 50$ GeV with $c^A_\mu / \Lambda = 0.1$ TeV$^{-1}$.}
\label{fig:3}
\end{figure*}

\begin{figure*}[ht!]
\centering{\includegraphics[width=0.48\textwidth]{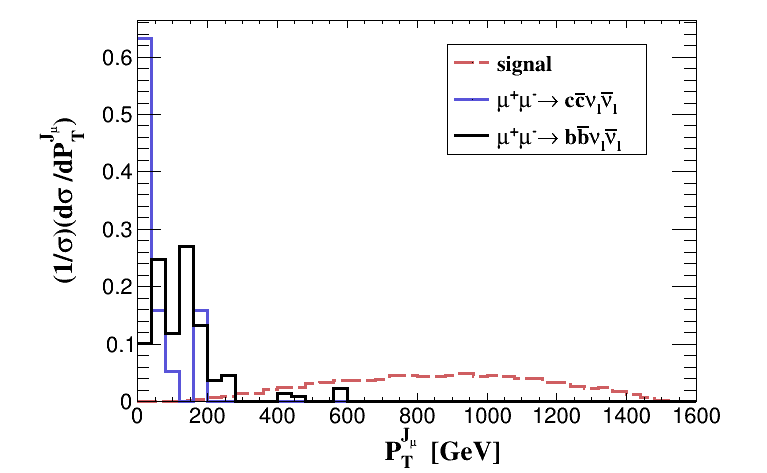}}
\centering{\includegraphics[width=0.48\textwidth]{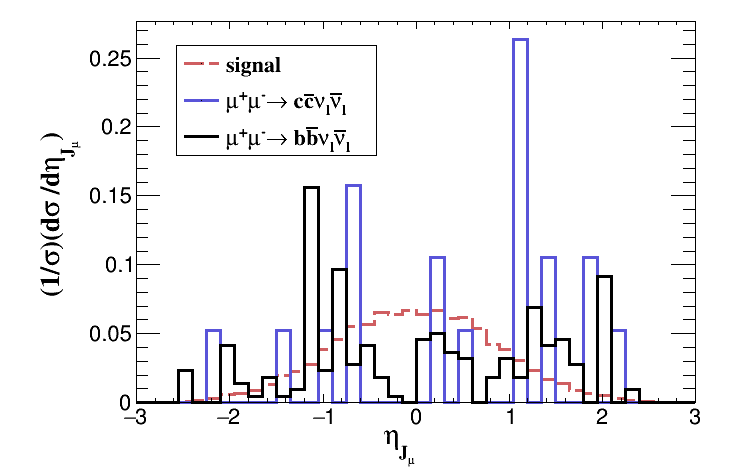}}
\centering{\includegraphics[width=0.48\textwidth]{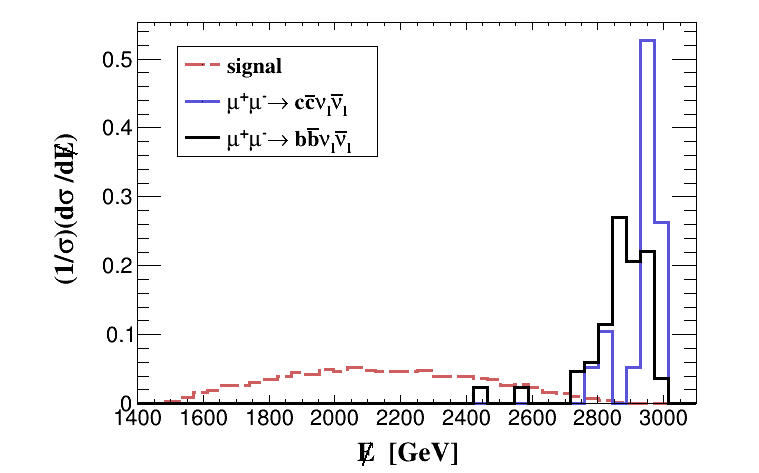}}
\centering{\includegraphics[width=0.48\textwidth]{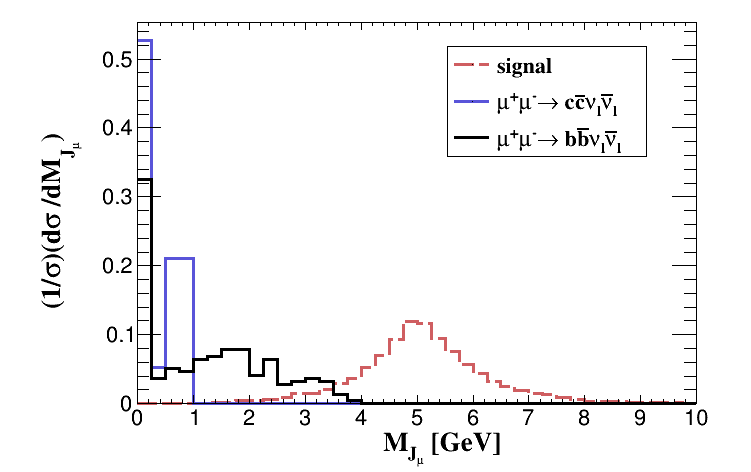}}
\caption{Some signal and background kinematic distributions for the signature of a $J_{\mu}$ plus ${\:/\!\!\!\! E}$ at a $\mu^+\mu^-$  collider, $P_T^{J_{\mu}}$, $\eta_{J_{\mu}}$,  ${\:/\!\!\!\! E}$  and $M_{J_{\mu}}$ distributions for $m_a = 5$ GeV with $c^A_\mu / \Lambda = 0.1$ TeV$^{-1}$.}
\label{fig:10}
\end{figure*}

\begin{figure*}[ht!]
\centering{\includegraphics[width=0.48\textwidth]{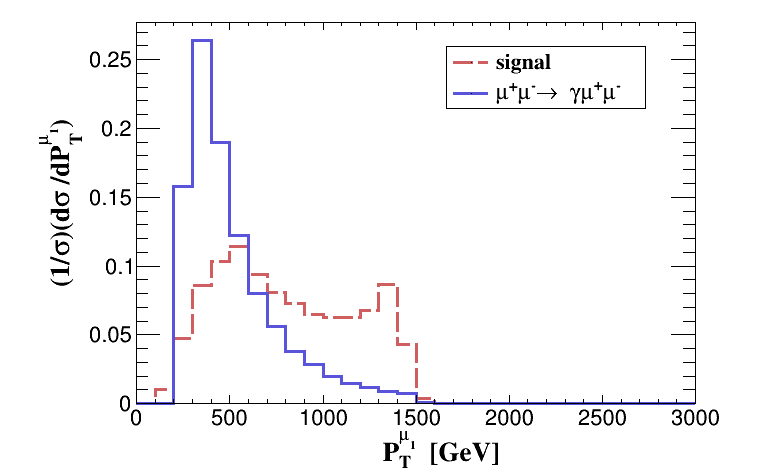}}
\centering{\includegraphics[width=0.48\textwidth]{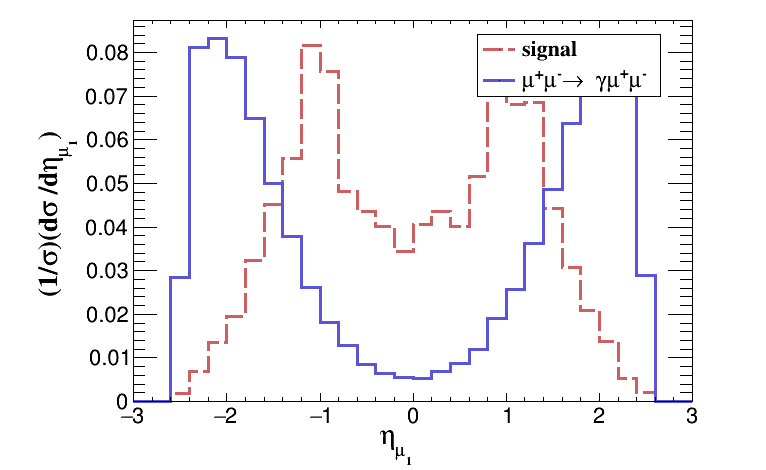}}
\centering{\includegraphics[width=0.48\textwidth]{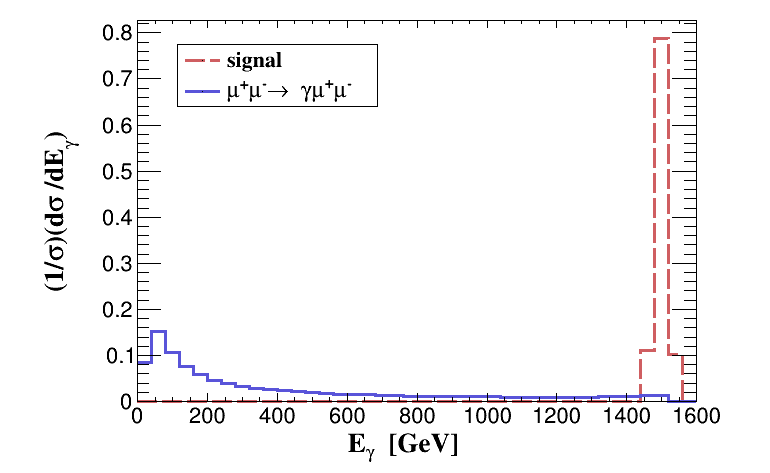}}
\centering{\includegraphics[width=0.48\textwidth]{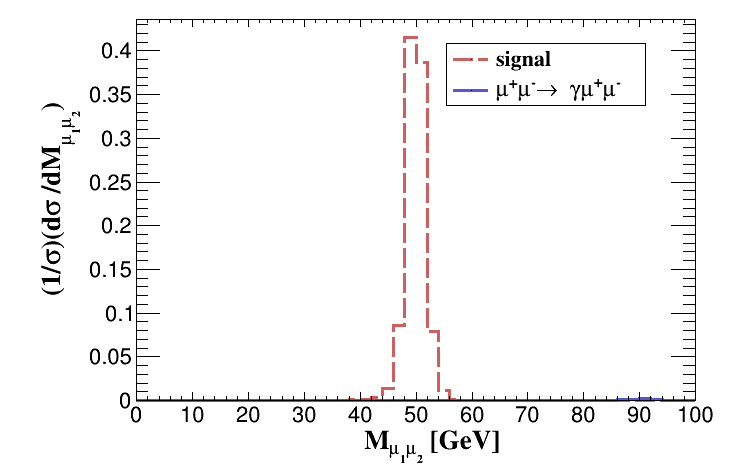}}
\centering{\includegraphics[width=0.48\textwidth]{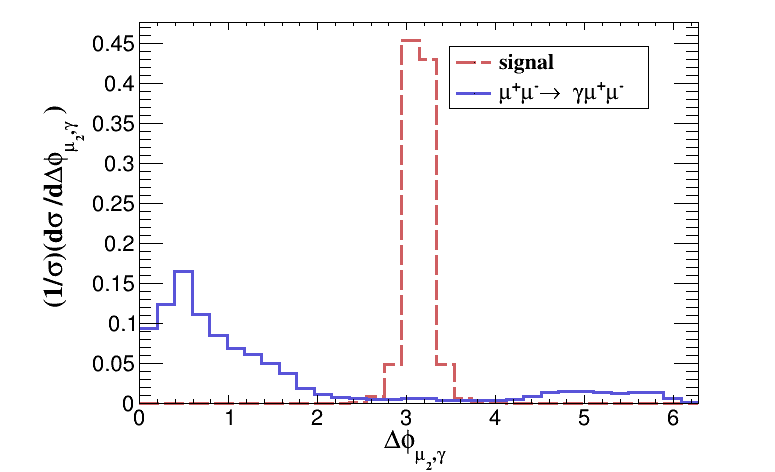}}
\centering{\includegraphics[width=0.48\textwidth]{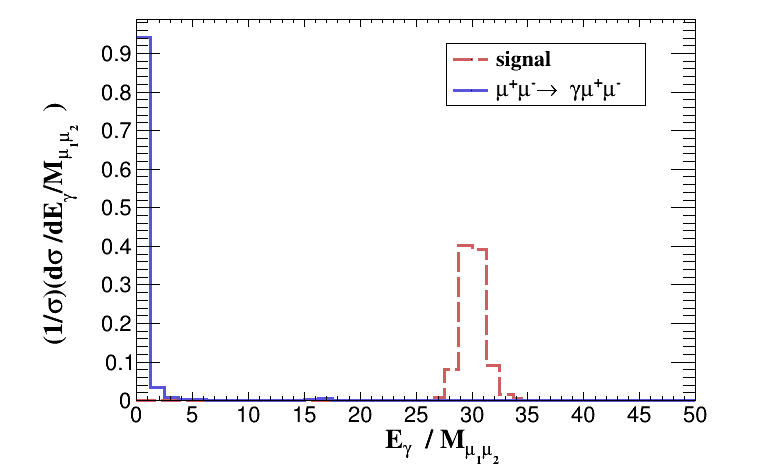}}
\caption{Some signal and background kinematic distributions for the signature of two isolated muons plus a $\gamma$ at a $\mu^+\mu^-$  collider, $P_T^{\mu_1}$, $\eta_{\mu_1}$, $E_{\gamma}$, $M_{\mu_1\mu_2}$, $\Delta\phi_{\mu_1,\gamma}$ and $E_{\gamma}/M_{\mu_1\mu_2}$ distributions for $m_a = 50$ GeV with $c^A_\mu / \Lambda = 10$ TeV$^{-1}$.}
\label{fig:auu}
\end{figure*}

\begin{figure*}[ht!]
\centering{\includegraphics[width=0.48\textwidth]{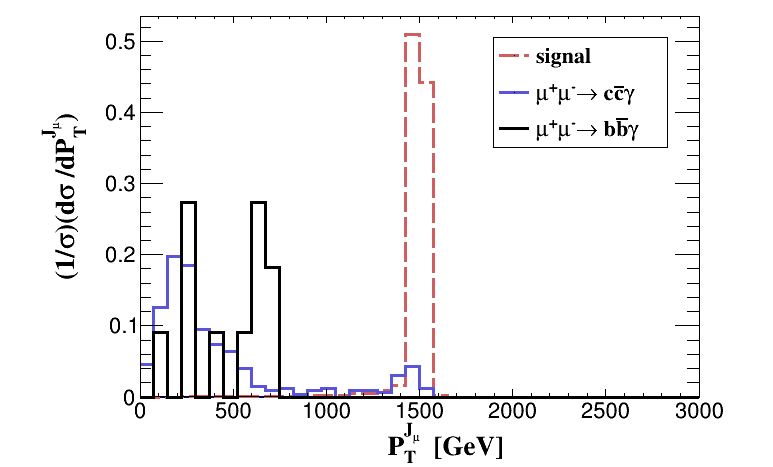}}
\centering{\includegraphics[width=0.48\textwidth]{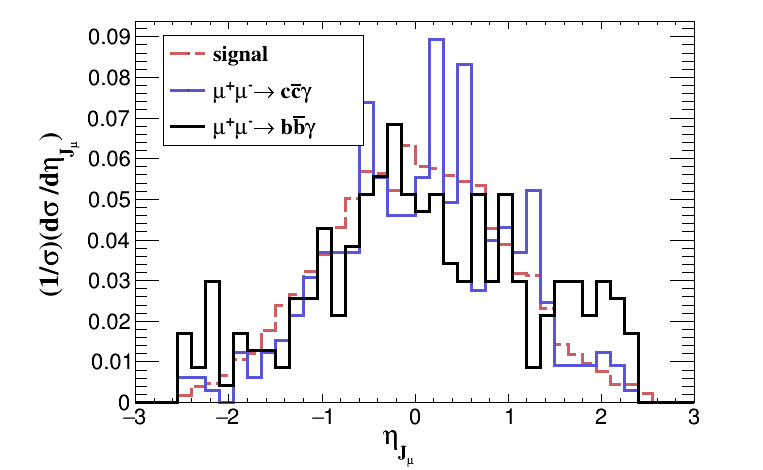}}
\centering{\includegraphics[width=0.48\textwidth]{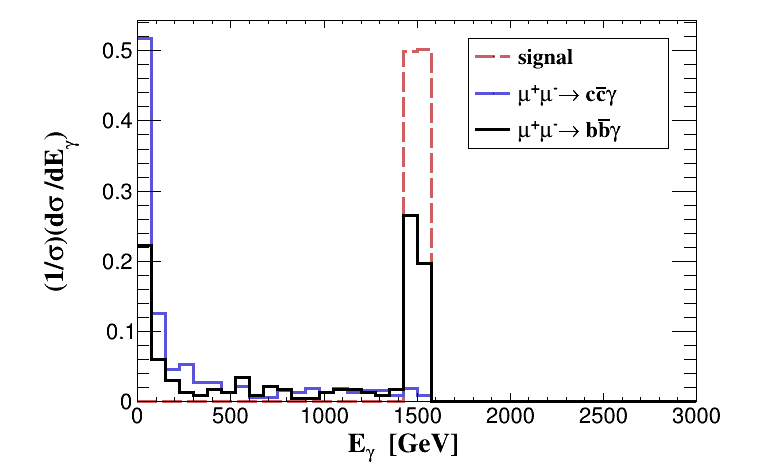}}
\centering{\includegraphics[width=0.48\textwidth]{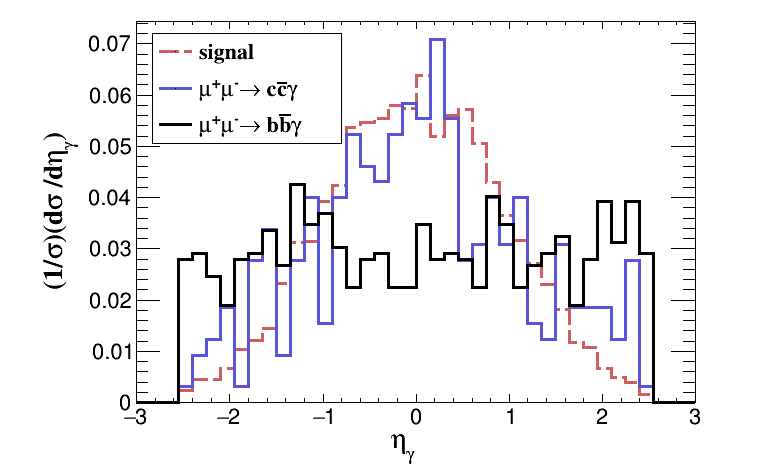}}
\centering{\includegraphics[width=0.48\textwidth]{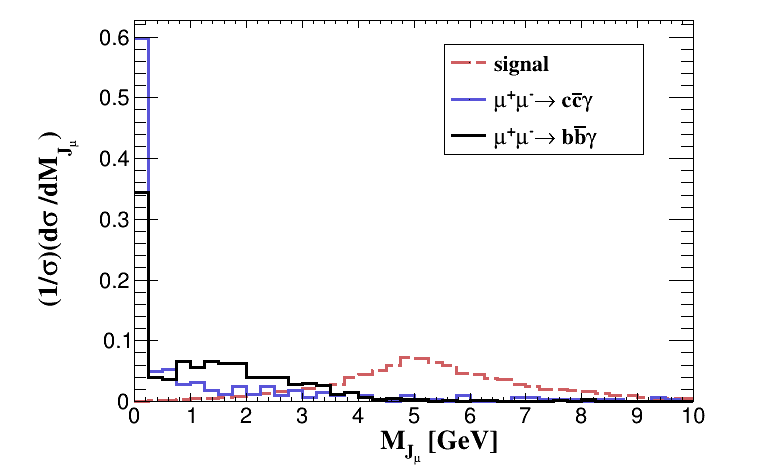}}
\centering{\includegraphics[width=0.48\textwidth]{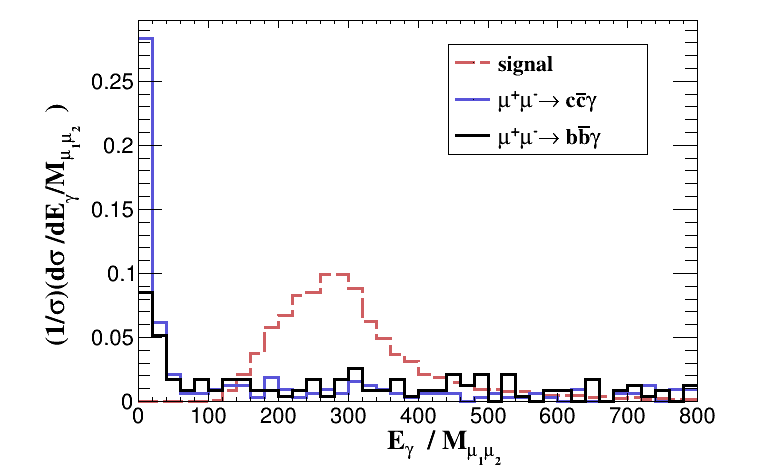}}

\caption{Some signal and background kinematic distributions for the signature of a $J_{\mu}$ plus a $\gamma$ at a $\mu^+\mu^-$  collider, $P_T^{J_{\mu}}$, $\eta_{J_{\mu}}$,  $E_{\gamma}$,  $\eta_{\gamma}$,  $M_{J_{\mu}}$ and $E_{\gamma}/M_{J_{\mu}}$distributions for $m_a = 5$ GeV with $c^A_\mu / \Lambda = 10$ TeV$^{-1}$.}
\label{fig:j10}
\end{figure*}

\begin{figure*}[ht!]
\centering{\includegraphics[width=0.44\textwidth]{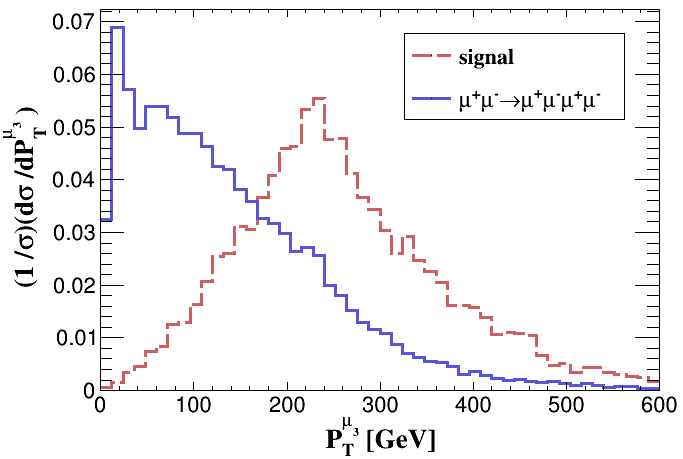}}
\hspace{0.55cm}
\centering{\includegraphics[width=0.44\textwidth]{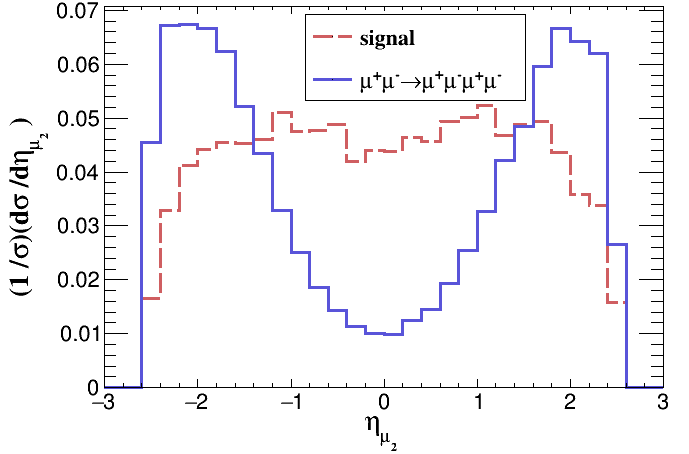}}
\centering{\includegraphics[width=0.44\textwidth]{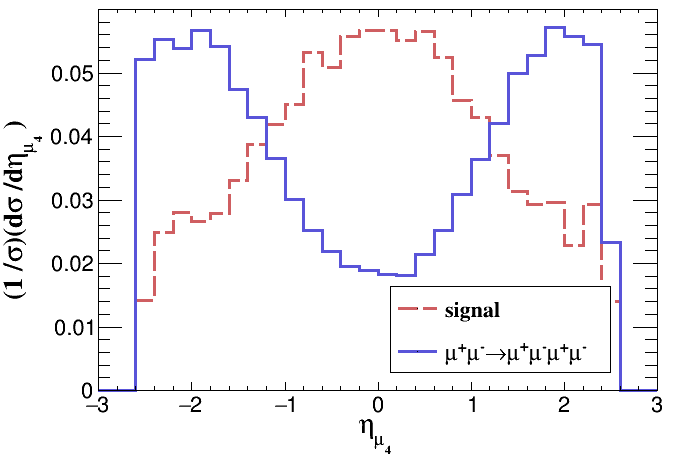}}
\hspace{0.55cm}
\centering{\includegraphics[width=0.44\textwidth]{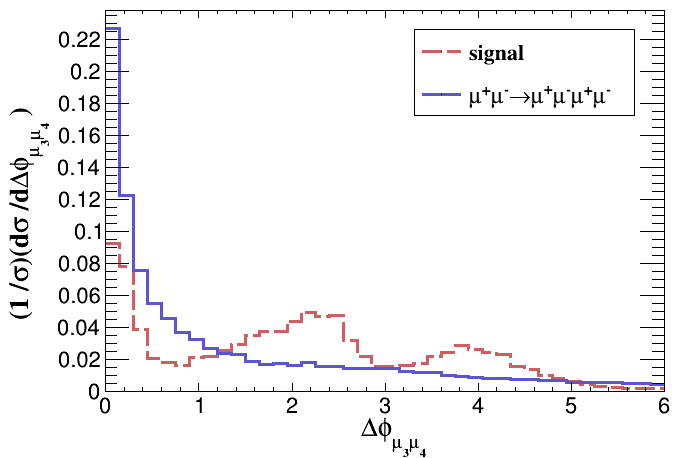}}
\centering{\includegraphics[width=0.44\textwidth]{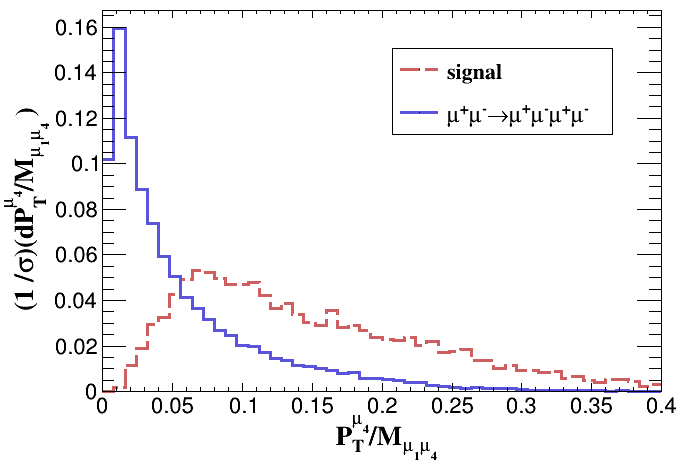}}
\hspace{0.55cm}
\centering{\includegraphics[width=0.44\textwidth]{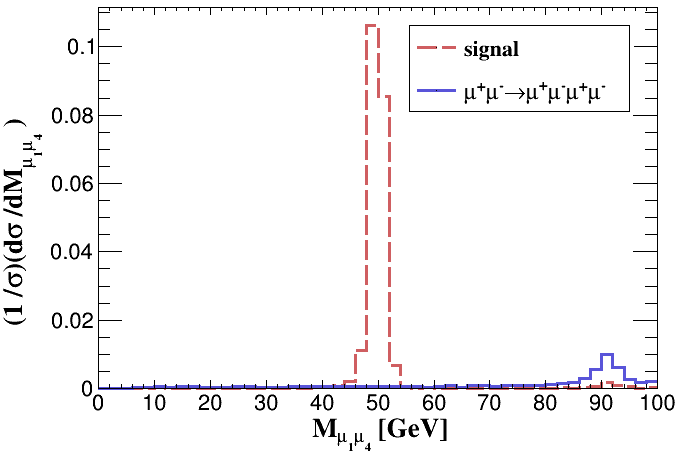}}
\caption{Some signal and background kinematic distributions for the signature of four isolated muons at a $\mu^+\mu^-$  collider, $P_T^{\mu_3}$, $\eta_{\mu_2}$, $\eta_{\mu_4}$ $\Delta\phi_{\mu_3\mu_4}$, $P_T^{\mu_4}/M_{\mu_{1}\mu_4}$ and $M_{\mu_{1}\mu_{4}}$ distributions for $m_a = 50$ GeV with $c^A_\mu / \Lambda = 10$ TeV$^{-1}$.}
\label{fig:zuu}
\end{figure*}

\begin{figure*}[ht!]
\centering{\includegraphics[width=0.44\textwidth]{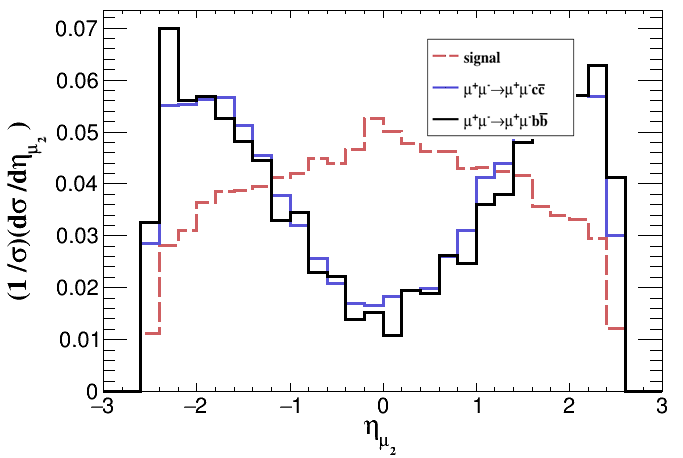}}
\hspace{0.55cm}
\centering{\includegraphics[width=0.44\textwidth]{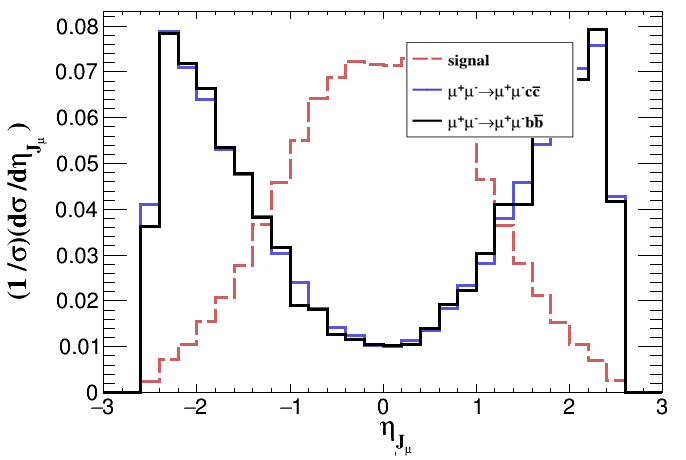}}
\centering{\includegraphics[width=0.44\textwidth]{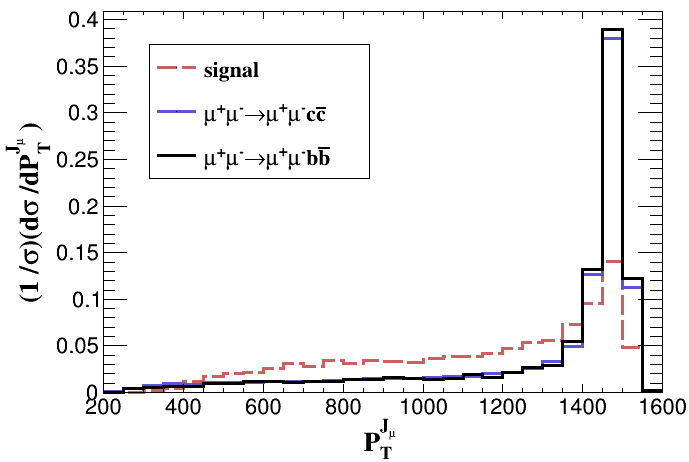}}
\hspace{0.5cm}
\centering{\includegraphics[width=0.45\textwidth]{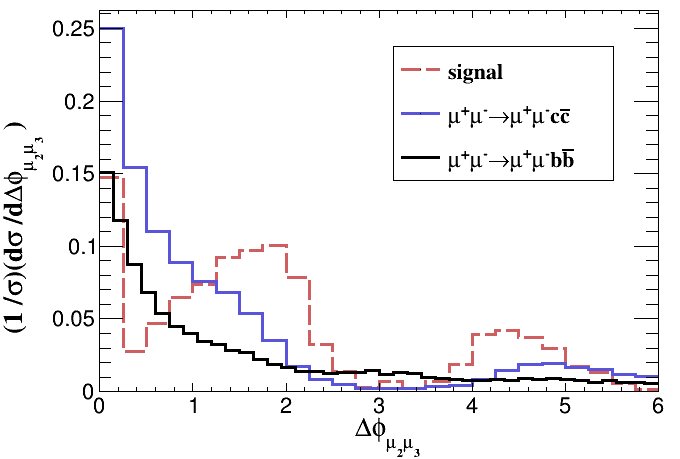}}
\centering{\includegraphics[width=0.44\textwidth]{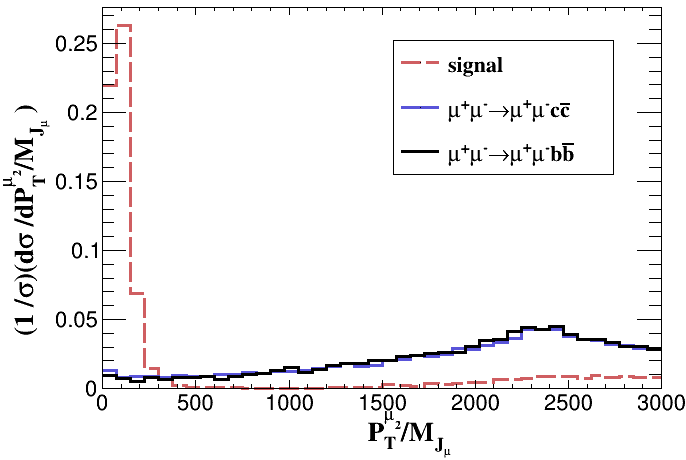}}
\hspace{0.43cm}
\centering{\includegraphics[width=0.45\textwidth]{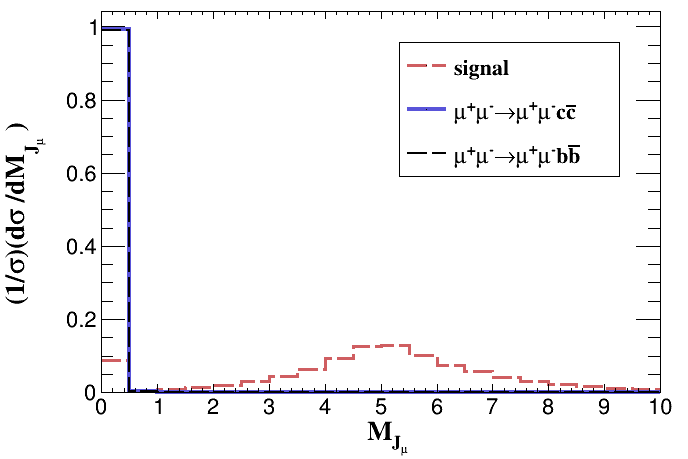}}
\caption{Some signal and background kinematic distributions for the signature of a $J_{\mu}$ plus two isolated muons at a $\mu^+\mu^-$  collider, $\eta_{\mu_2}$, $\eta_{J_{\mu}}$, $P_T^{J_{\mu}}$, $\Delta \phi_{\mu_{2}\mu_3}$, $M_{\mu_{1}\mu_4}/(P_T^{\mu_1}+P_T^{\mu_4})$ and $M_{J_{\mu}}$ distributions for $m_a = 5$ GeV with $c^A_\mu / \Lambda = 10$ TeV$^{-1}$.}
\label{fig:zuu_jet_all}
\end{figure*}

In this Appendix, we choose some representative kinematic distributions for both signals and backgrounds at at a muon collider in the following : 
\begin{itemize}
\item For the signature of two isolated muons plus ${\:/\!\!\!\! E}$ at at a $\mu^+\mu^-$  collider, $P_T^{\mu_1}$, $\eta_{\mu_1}$, ${\:/\!\!\!\! E}$, $M_{\mu_1\mu_2}$, $\Delta\phi_{\mu_1,{\:/\!\!\!\! E}}$ and $\Delta\phi_{\mu_2,{\:/\!\!\!\! E}}$ distributions for $m_a = 50$ GeV with $c^A_\mu / \Lambda = 0.1$ TeV$^{-1}$ are shown in Fig.~\ref{fig:3}.
\item For the signature of a $J_{\mu}$ plus ${\:/\!\!\!\! E}$ at a $\mu^+\mu^-$  collider, $P_T^{J_{\mu}}$, $\eta_{J_{\mu}}$,  ${\:/\!\!\!\! E}$  and $M_{J_{\mu}}$ distributions for $m_a = 5$ GeV with $c^A_\mu / \Lambda = 0.1$ TeV$^{-1}$ are shown in Fig.~\ref{fig:10}.
\item For the signature of two isolated muons plus a $\gamma$ at at $\mu^+\mu^-$  colliders, $P_T^{\mu_1}$, $\eta_{\mu_1}$, $E_{\gamma}$, $M_{\mu_1\mu_2}$, $\Delta\phi_{\mu_1,\gamma}$ and $E_{\gamma}/M_{\mu_1\mu_2}$ distributions for $m_a = 50$ GeV with $c^A_\mu / \Lambda = 10$ TeV$^{-1}$ are shown in Fig.~\ref{fig:auu}.
\item For the signature of a $J_{\mu}$ plus a $\gamma$ at a $\mu^+\mu^-$  collider, $P_T^{J_{\mu}}$, $\eta_{J_{\mu}}$,  $E_{\gamma}$,  $\eta_{\gamma}$,  $M_{J_{\mu}}$ and $E_{\gamma}/M_{J_{\mu}}$ distributions for $m_a = 5$ GeV with $c^A_\mu / \Lambda = 10$ TeV$^{-1}$ are shown in Fig.~\ref{fig:j10}.
\item For the signature of four isolated muons at a $\mu^+\mu^-$  collider, $P_T^{\mu_3}$, $\eta_{\mu_2}$, $\eta_{\mu_4}$ $\Delta\phi_{\mu_3,\mu_4}$, $P_T^{\mu_4}/M_{\mu_{1}\mu_4}$ and $M_{\mu_{1}\mu_{4}}$ distributions for $m_a = 50$ GeV with $c^A_\mu / \Lambda = 10$ TeV$^{-1}$ are shown in Fig.~\ref{fig:zuu}.
\item For the signaturte of two isolated muons plus a $J_{\mu}$ at a $\mu^+\mu^-$  collider, $\eta_{\mu_2}$, $\eta_{J_{\mu}}$, $P_T^{J_{\mu}}$, $\Delta \phi_{\mu_{2}\mu_3}$, $M_{\mu_{1}\mu_4}/(P_T^{\mu_1}+P_T^{\mu_4})$ and $M_{J_{\mu}}$ distributions for $m_a = 5$ GeV with $c^A_\mu / \Lambda = 10$ TeV$^{-1}$ are shown in Fig.~\ref{fig:zuu_jet_all}.
\end{itemize}  

 On the other hand, we modify the event selections for detecting two isolated muons plus ${\:/\!\!\!\! E}$ with varying $m_a$ at a muon collider (as shown in Table.~\ref{tab:other}) to optimize the signal efficiency. Specifically, we adjust the ranges of $\eta_{\mu_{1,2}}$ and $\eta_{\:/\!\!\!\! E}$ for small values of $m_a$, as loosening these criteria can improve signal detection while still eliminating all background events with the current selection criteria. In addition, we adjust ${\:/\!\!\!\! E}/M_{\mu_1\mu_2}$ based on signal and background distributions since it decreases as $m_a$ increases. Conversely, we do not optimize event selections for detecting a $J_{\mu}$ plus ${\:/\!\!\!\! E}$ with varying $m_a$ at a muon collider since the relevant backgrounds are already unlikely to satisfy the conditions of two detectable muons in the muon spectrometer and forming an energetic $J_{\mu}$ in the central region. As a result, nearly all of these signals are free from background events after the cut-(3) selection in Table.~\ref{tab:CF_uu_2}. 
 Similarly, we fine-tune event selections for detecting two isolated muons plus a photon and four isolated muons at a muon collider for different $m_a$ in the \textbf{EWP} scenario, as listed in Table~\ref{tab:auu_tune} and Table~\ref{tab:zuu_tune}.

\begin{table}[htp]
	\begin{center}\begin{tabular}{|c|c|c|c|c|c|}\hline $m_a$ [GeV] & $\lvert\eta_{\mu_{1,2}}\rvert $ & $\lvert\eta_{{\:/\!\!\!\! E}}\rvert $ & ${\:/\!\!\!\! E}/M_{\mu_1\mu_2}$ & $\Delta M_{\mu_1\mu_2}$ & $\Delta\phi_{\mu_2, {\:/\!\!\!\! E}}$  \\
			\hline $10$ &  $< 3.0$  &  $< 3.0$  & $> 140$ &  same  &  $(2.5,3.6)$ \\ 
			\hline $20$ &  $< 2.0$  &  $< 2.0$  &  $> 70$  &  same  &  $(2.7,3.6)$ \\ 
			\hline $30$ &  $< 1.8$  &  $<1.9$  &  $> 50$  &  same  &  same \\ 
			\hline $40$ &  $< 1.6$  &  same  &  $> 40$  &  same  &  same \\ 
			\hline $60$ &  same  &  same  &  $> 22$  &  $< 1.5$  &  same \\ 
			\hline $70$ &  same  &  same  &  $> 22$  &  $< 1.5$  &  same \\ 
			\hline $80$ &  same  &  same  &  $> 20$  &  $< 2.3$  &  $(2.8,3.5)$ \\ 
			\hline $85$ &  same  &  same  &  $> 18$  &  $< 2.3$  &  $(2.8,3.5)$ \\
			\hline \end{tabular} \caption{The changes of some event selections in the \textbf{EWV} scenario with some $m_a$ benchmark points for two isolated muons plus ${\:/\!\!\!\! E}$ at a muon collider where $\Delta M_{\mu_1\mu_2}\equiv \lvert M_{\mu_1\mu_2}-m_a\rvert$ and "same" means the same event selection as the benchmark point $m_a = 50$ GeV in the main text.}
		\label{tab:other}
	\end{center}
\end{table}

\begin{table}[htp]
	\begin{center}\begin{tabular}{|c|c|c|c|c|c|}\hline $m_a$ [GeV] & $P_T^{\mu_2} $ & $\lvert\eta_{\mu_{1,2}}\rvert $ & $\lvert\eta_{E_{\gamma}}\rvert $ & $E_{\gamma}/M_{\mu_1\mu_2}$ & $\Delta M_{\mu_1\mu_2}$  \\
			\hline $30$ &  $> 300$  &  same  &  $< 1.8$  &  $> 46$  &  same \\ 
			\hline $40$ &  $> 400$  &  same  &  same  &  $> 35$  &  $< 1.8$ \\ 
			\hline $60$ &  same  &  $< 1.2$  &  $< 1.2$  &  $> 24$  &  $< 2.2$ \\ 
			\hline $70$ &  same  &  $< 1.2$  &  $< 1.2$  &  $(20,23)$  &  $< 2.8$ \\ 
			\hline $80$ &  same  &  $< 1.2$  &  $< 1.2$  &  $(17,20)$  &  $< 3.0$ \\ 
			\hline $85$ &  same  &  $< 1.2$  &  $< 1.2$  &  $(16,19)$  &  $< 3.0$ \\
			\hline \end{tabular} \caption{Similar to Table.~\ref{tab:other}, but for the $\mu^{+}\mu^{-}\rightarrow \gamma a$ channel.}
		\label{tab:auu_tune}
	\end{center}
\end{table}

\begin{table}[htp]
	\begin{center}\begin{tabular}{|c|c|c|c|c|c|}\hline $m_a$ [GeV] & $\lvert\eta_{\mu_{1}}\rvert $ & $\lvert\eta_{\mu_{4}}\rvert $ & $\Delta \phi_{\mu_{3},\mu_4}$ & $P_{T}^{\mu_4}/M_{\mu_{1}\mu_{4}} $ & $\Delta M_{\mu_1\mu_4}$  \\
			\hline $30$ &  same   &  $<1.6$   &  $>0.8$ &  same  &  $<4.0$ \\ 
			\hline $40$ &  same   &  same     &  same   &  same  &  $<4.0$ \\ 
			\hline $60$ &  same   &  same     &  same   &  $>0.06$   &  same \\ 
			\hline $70$ &  $<1.8$  &  $< 1.4$  &  $>1.2$  &  $>0.06$  &  same \\ 
			\hline $80$ &  $<1.8$  &  $< 1.4$  &  $>1.2$  &  $>0.08$  &  same \\ 
			\hline $85$ &  $<1.8$  &  $< 1.4$  &  $>1.2$  &  $>0.08$  &  $< 4.0$ \\
			\hline \end{tabular} \caption{Similar to Table.~\ref{tab:other}, but for the $\mu^{+}\mu^{-}\rightarrow \mu^{+}\mu^{-} a$ channel. }
		\label{tab:zuu_tune}
	\end{center}
\end{table}

\end{document}